\documentclass[epj]{svjour}
\RequirePackage{graphicx}
\RequirePackage[colorlinks,citecolor=blue,urlcolor=blue,linkcolor=blue]{hyperref}
\RequirePackage{lineno}

\usepackage{tabularx}
\usepackage{cite}
\usepackage{booktabs}
\usepackage{amsmath}
\usepackage{graphicx}% Include figure files
\usepackage{dcolumn}% Align table columns on decimal point
\usepackage{bm}% bold math
\usepackage{hyperref}
\usepackage{nicefrac} 
\usepackage{xfrac}
\newenvironment{color}[3]
{
}{
}

\newcommand{\grey}[1]     {}

\newcommand{\pT}{$p_{\rm{T}}$}
\newcommand{\mT}{$m_{\rm{T}}$}

\begin{document}

\title{Centrality, transverse momentum and collision energy dependence of the Tsallis parameters in relativistic heavy-ion collisions}

\author{Rajendra Nath Patra\inst{1, 2, 3}\thanks{e-mail:  rajendra.nath.patra@cern.ch,\\rajendrapatra07@gmail.com}
	\and 	Bedangadas Mohanty\inst{4}
     \and   Tapan K. Nayak\inst{4, 2}}
 
\institute{Bogolyubov Institute for Theoretical Physics, National Academy of Sciences of Ukraine, Kiev, Ukraine \label{addr1} 
          \and      CERN, CH 1211, Geneva 23, Switzerland \label{addr2} 
           \and      University of Jammu, Jammu 180006, India \label{addr3} 
          \and     National Institute of Science Education and Research,  Jatni 752050,
India \label{addr4}
}
\date{Received: date / Accepted: date}

%===========================ABSTRACT==========================%

\abstract{The thermodynamic properties of matter created in high-energy
heavy-ion collisions have been studied in the framework of the
non-extensive Tsallis statistics. The transverse momentum 
($p_{\rm  T}$)~spectra of identified charged particles (pions, kaons,
protons) and all charged particles from the available experimental
data of Au-Au collisions at the Relativistic Heavy Ion Collider (RHIC)
energies and Pb-Pb collisions at the Large Hadron Collider (LHC)
energies are fitted by the Tsallis distribution. The fit parameters, 
$q$ and $T$ measure the degree of deviation from an equilibrium state and 
the effective temperature of the thermalized system, respectively. 
The $p_{\rm  T}$~spectra are well described by the Tsallis
distribution function from peripheral to central collisions 
for the wide range of collision energies, from $\sqrt{s_{\rm NN}}$ = 7.7
GeV to 5.02 TeV.
The extracted Tsallis parameters are found to be 
dependent on the particle species, collision energy, centrality, and
fitting ranges in $p_{\rm T}$.
For central collisions, both $q$ and $T$ depend strongly 
on the fit ranges in $p_{\rm T}$. 
For most of the collision energies,
$q$ remains almost constant as a function of centrality,
whereas $T$ increases from peripheral to central collisions.
For a given centrality, $q$ systematically increases as a function of
collision energy whereas $T$ has a decreasing trend. A profile plot of $q$ and $T$ with respect to collision energy and centrality shows an anti-correlation between the two parameters. }
%A quantitative description of the Tsallis parameters with
%different fitting ranges of the $p_{\rm T}$ is presented as a function
%of collision energy and centrality. 

%\PACS{25.75.-q,25.75.Nq,12.38.Mh}
%\keywords{Tsallis distribution, heavy-ion, kinetic temperature, transverse momentum, collision energy}

%\pacs{25.75.-q,25.75.Nq,12.38.Mh}
\maketitle
\twocolumn
%============================MAIN==========================%
\section{Introduction}\label{intro}
Relativistic heavy-ion collisions produce matter at extreme conditions
of temperatures and energy densities in the form of quark-gluon plasma
(QGP). The QGP state is formed at the early stages of the collision, which
survives for a very short span of time ($\sim$7--10~fm/$c$), after
which the matter gets transformed rapidly to a system of hadron
gas. The information about the initial condition of the system gets
mostly lost by multi-partonic interactions throughout the evolution
of the collision. The final state behavior of such a colliding
system can be obtained from the measurement of the number and identity
of produced particles, along with their energy and momentum
spectra. The final state information is nevertheless most useful for
understanding the particle production mechanisms and the nature of the
matter produced in these high-energy collisions.

The space-time evolution of the hot and dense system produced in the
collision ceases at two symbolic freeze-out conditions, namely,
the chemical freeze-out and kinetic freeze-out. 
The colliding medium first reaches chemical equilibrium, and then cools down by expansion. Due to this expansion of the system, the inelastic collisions cease when the mean free path for the interactions becomes comparable to the system size. This is known as the chemical freeze-out, at which point the abundances of different particle species become constant.
After this
stage, even if the relative fractions of the particles are constant,
these particles continue to interact till a point where
the final state interactions between the hadrons are no longer
effective. After this kinetic freeze-out all the interactions cease
 and the (transverse) momentum spectra of the particles remain
 unchanged. Therefore, transverse momentum ($p_{\rm T}$) spectra of
 produced particles constitute some of the basic measurements to extract the kinetic freeze-out condition of the systems produced in
 high-energy collisions.

Several different standard statistical model fits using Boltzmann-Gibbs (BG), Fermi-Dirac, Bose-Einstein distributions, \mT-exponential distribution~\cite{STAR-2009, STAR-BES1} Erlang distribution~\cite{Gao-2012_Erlang}, Tsallis distribution~\cite{Wong-2015, Azmi-2020, Cleymans-Worku-2012-EPJ, Trambak-2018, Azmi-2015} and other kinds of distribution functions have been used to describe the $p_{\rm T}$ spectra and extract physical parameters.
The BG statistical model has attained a huge success
in explaining the thermodynamic properties in different physical
systems. However, the BG thermal (exponential) model is inappropriate at high transverse momenta
in high-energy collision systems as it cannot explain the high transverse
momentum range which is mostly followed by the inverse power-law
behavior~\cite{NA61-1982, Hagedron-1983, ALICE-pp_2013}. A distribution following the non-extensive Tsallis
statistics~\cite{Tsallis-1988,Tsallis-1998} has been shown to provide
a better description of the nature of the non-exponential transverse
momentum spectrum. The description of particle spectra using the
Tsallis distribution has gained lots of interest recently as it can
describe the spectra at high $p_{\rm T}$. The Tsallis distribution is
found to be very successful
in describing the spectra at high $p_{\rm T}$ ranges in proton-proton ($pp$)
collisions~\cite{Azmi-2015, Wong-2015}. The $p_{\rm T}$ spectra of
identified particles in $pp$ system at different collision energies can
be explained by the Tsallis distribution function as presented in
Ref.~\cite{Trambak-2018, Zheng-2016}, whereas the multiplicity
dependence of all charged particles spectra with the Tsallis framework
is reported in Ref.~\cite{Rath-2020}. 

In case of heavy-ion collisions, the transverse momentum spectra have traditionally been fitted by
blast-wave model~\cite{BW1}~\cite{STAR-BES1,STAR-14.5,
  STAR-2009,ALICE-2.76-2013,ALICE-5.02-2020}. This model has a simple
assumption that there is a kinetic freeze-out of the colliding medium
at temperature  $T_{\rm kin}$ and particles are moving with a common
collective radial flow velocity ($\beta$). However, the blast-wave
model cannot be used to describe the spectra at mid to high-$p_{\rm T}$
ranges. Recent studies \cite{Azmi-2020} have found that
the Tsallis statistics also works in
heavy-ion collisions providing good fitting of the $p_{\rm T}$
spectra of all charged particles.
Since these fits do not account for the radial flow,
the temperatures obtained from the Tsallis fits are the effective 
temperatures, different from the kinetic freeze-out temperatures. 
This effective 
temperature depends on the particle mass and the collective
flow velocity. The intercept in $T=T_{\rm kin} +
\sfrac{1}{2}~m\beta_{\rm T}^2$ is an alternative
method~\cite{PHENIX-2004, Wei-2016, Lao-2018} to find freeze-out
temperature where $m$ is the rest mass and $\beta_{\rm T}$ is the
average transverse radial flow velocity. 

In the present work, we use the thermodynamically consistent form of the
Tsallis distribution~\cite{Cleymans-Worku-2012-EPJ} to fit the
transverse momentum ($p_{\rm T}$) spectra of the identified charged
particles with a focus on pions as well as all charged particles in
heavy-ion collision systems corresponding to eight different collision
energies at RHIC and two collision energies at the LHC. 
For each of the spectrum, we obtain the Tsallis fit parameters, $q$,
which expresses the degree of deviation from an equilibrium state, and
$T$ as the effective temperature. 
The dependencies of the fitting
parameters on collision centrality, collision energy and fitting
ranges in $p_{\rm T}$ are investigated and will be presented. The
relative success of Tsallis distribution for different particles in
heavy-ion system will be also discussed. In section~\ref{methodology},
we discuss the formulation of the Tsallis distribution and the methodology
of fitting the  $p_{\rm T}$ spectra. In
section~\ref{pT_spectrum_fitting}, we present Tsallis fittings of the
$p_{\rm T}$ spectra at different centrality, energy and $p_{\rm T}$
ranges. In section~\ref{results}, present the Tsallis fit parameters
and discuss their relevances for different collision energies. A
discussion on the results is given in
section~\ref{discussion}. Finally, the paper is summarized in
section~\ref{summary}.

\section{Tsallis statistics and methodology}\label{methodology}
The non-extensive Tsallis form of the Boltzmann-Gibbs distribution in terms of energy ($E$), effective temperature ($T$), and chemical potential ($\mu$) of a system can be expressed as~\cite{Tsallis-1988,Tsallis-1998},
\begin{equation} \label{Tsallis:BG}
f(E,q,T,\mu) \equiv \exp_{\rm q} \left(- \frac{E-\mu}{T}\right) \equiv \left( 1+ ( q-1) \frac{E-\mu}{T} \right)^{-\frac{1}{q-1}}.
\end{equation} 
The $\exp_{\rm q}(x)$ has the form,
\begin{equation} \label{expq}
\exp_{\rm q}(x)\equiv \left\{
\begin{array}{ll}
\left(1+(q-1)x \right)^\frac{1}{q-1}  & \mbox{if $x>0$},\\
\left(1+(1-q)x \right)^\frac{1}{1-q}  & \mbox{if $x\leq0$},
\end{array}
\right. 
\end{equation}
where $q$ is the entropy index which measures the degree of non-additivity of the entropy or deviation from the equilibrium of the system. In general, $q \geq 1$.  In equilibrium case, \textit{i.e.}, in the limit $q \rightarrow 1, \exp_{\rm q}(x) \rightarrow \exp(x)$. Therefore, Eq.~\ref{Tsallis:BG} is simplified to the extensive Boltzmann-Gibbs form, 
\begin{equation}  \label{eq-BG}
f_{ q \rightarrow 1}(E, q, T \mu)\equiv  f_{\rm BG}(E,T,\mu) \equiv \exp\left(- \frac{E-\mu}{T}\right).
\end{equation}
The Tsallis form with thermodynamical consideration of the invariant
momentum distribution of the particles has been introduced and
discussed in
\cite{Cleymans-Worku-2012-JPG,Cleymans-Worku-2012-EPJ}. The particle number
density ($n$) can be expressed using Tsallis statistics as, 
\begin{equation} \label{Tsallis:number}
n=g \int\frac{d^3p}{(2\pi)^3} f^q,
\end{equation}
where $g$ is the degeneracy factor of the particles.
The expression of the invariant momentum distribution can be obtained from the particle number density as given in Eq.~\ref{Tsallis:number} as,
\begin{equation} \label{Tsallis:invariant}
E\frac{d^3N}{dp^3} = \frac{gV}{(2\pi)^3}E \left( 1+ ( q-1)\frac{E-\mu}{T} \right)^{-\frac{q}{q-1}}.
\end{equation}
Here $N$ is the particle number, and $V$ is refereed as the volume of the system which depends on the $q$. Therefore, $V$ is not necessarily the volume of the system, but serves as a normalization factor of the Tsallis distribution. For non vanishing chemical potential, Eq.~\ref{Tsallis:invariant} can be rewritten in terms of transverse mass ($m_{\rm T}$) and rapidity ($y$) in the form,
\begin{equation} \label{Tsallis:rapidity}
\begin{split}
\frac{1}{2\pi p_{\rm T}} \frac{d^2N}{dp_{\rm T} dy} =  \frac{gV}{(2\pi)^3}  m_{\rm T} \cosh(y) \\ 
 \left( 1+ ( q-1)\frac{m_{\rm T} \cosh(y) -\mu}{T} \right)^{-\frac{q}{q-1}}.
\end{split}
\end{equation}
The invariant yields, $\frac{1}{2\pi p_{\rm T}} \frac{d^2N}{dp_{\rm T} dy}$ is an experimentally measured observable. Therefore, the Tsallis parameters $q,~T$ can be obtained from experimentally measured invariant momentum distributions using the fitting function as given in Eq.~\ref{Tsallis:rapidity}, with $\mu$ as an input.
%%%%%%%%%%%%%%%%%%%%%%%%%%%%%%%%%%%%%%%%%%%%

\begin{table*} [htbp!]
	\caption{Collision energy, collision species, observed 
		particles, centrality binning, pseudo-rapidity, and 
		transverse momentum range of \pT~spectra from the STAR, PHENIX and ALICE experiments.}
	\centering 
	\begin{tabular}{ >{\centering}p{1.6cm}  >{\centering}p{1.2cm}  >{\centering}p{2.5cm}   >{\centering}p{1.0cm}    >{\centering}m{5.2cm}  l  } \toprule
%		\begin{tabular}{ p{1.8cm}   p{1.5cm}  c  p{1.0cm}   p{1.5cm}   m{5.2cm}  l  } \toprule
	
		$\sqrt{s_{NN}}$ (GeV) & Collision system & Particles & $|y| (|\eta|)$  & Centrality (\%) & Experiment\\
		\toprule 
		7.7, 11.5, 19.6, 27, 39  	 &  Au-Au     &  $\pi^+, \pi^-$, ${\rm K^+, K^-}$, ${\rm p, \bar{p}}$ & 0.1  &  0-5, 5-10, 10-20, 20-30, 30-40, 40-50, 50-60, 60-70, 70-80   & STAR~\cite{STAR-BES1}      \\
		\hline 
		14.5 &  Au-Au    &      $\pi^+, \pi^-$, ${\rm K^+, K^-}$, ${\rm p, \bar{p}}$     & 0.1   & 0-5, 5-10, 10-20, 20-30, 30-40, 40-50, 50-60, 60-70, 70-80       & STAR~\cite{STAR-14.5}       \\
		\hline 
		62.4 &  Au-Au        &      $\pi^+, \pi^-$, ${\rm p, \bar{p}}$     & 0.5   & 0-10, 10-20, 20-40, 40-80       & STAR~\cite{STAR-62.4}       \\
		\hline 
		200 &  Au-Au        &       $\pi^+, \pi^-$, ${\rm p, \bar{p}}$     & 0.5   & 0-12, 10-20, 20-40, 40-60, 60-80       & STAR~\cite{STAR-200}       \\
		\hline 
		200 &  Au-Au        &    $\pi^+, \pi^-$, ${\rm K^+, K^-}$, ${\rm p, \bar{p}}$     & 0.26   & 0-5, 5-10, 10-15, 15-20, 20-30, 30-40, 40-50, 50-60, 60-70, 70-80, 80-92       & PHENIX~\cite{PHENIX-200}     \\
		\hline 
		2760 &  Pb-Pb        &       $\pi^+, \pi^-$, ${\rm K^+, K^-}$, ${\rm p, \bar{p}}$    & 0.5   & 0-5, 5-10, 10-20, 20-30, 30-40, 40-50, 50-60, 60-70, 70-80, 80-90       & ALICE~\cite{ALICE-2.76-2013}      \\
		\hline 
		2760 &  Pb-Pb        &    $\pi^+ + \pi^-$, ${\rm K^+ + K^-}$, ${\rm p + \bar{p}}$    & 0.5   & 0-5, 5-10, 10-20, 20-30, 30-40, 20-40, 40-50, 40-60, 60-80       & ALICE~\cite{ALICE-2.76-2016}      \\
		\hline 
		2760 &  Pb-Pb        &       N$_{\rm ch}$     & 0.8 & 0-5, 5-10, 10-20, 20-30, 30-40, 40-50, 50-60, 60-70, 70-80      & ALICE~\cite{ALICE-2.76-Nch}      \\
		\hline 
		5020 &  Pb-Pb        &      $\pi^+ + \pi^-$, ${\rm K^+ + K^-}$, ${\rm p + \bar{p}}$    & 0.5 &  0-5, 5-10, 10-20, 20-30, 30-40, 40-50, 50-60, 60-70, 70-80,  80-90       & ALICE~\cite{ALICE-5.02-2020}     \\
		\hline 
		5020 &  Pb-Pb        &       N$_{\rm ch}$    & 0.8 & 0-5, 5-10, 10-20, 20-30, 30-40, 40-50, 50-60, 60-70, 70-80        & ALICE~\cite{ALICE-5.02-Nch}      \\
		\toprule 
	\end{tabular}
	\label{Summary: all_Energy}
\end{table*}
%
%
%%%%%%%%%%%%%%%%%%%%%%%%%%%%%%%%%%%%%%%%%%%%%%%%%%%%%%%%%%%%%%%%%%%%%%%%%%%%%%%%%%%%%%%%

A different type of Tsallis distribution is used by the STAR~\cite{STAR-pp-2007}, PHENIX~\cite{PHENIX-pp-2010,PHENIX-pp-2011}, ATLAS~\cite{ATLAS-pp-2011}, CMS~\cite{CMS-pp-2017} and ALICE~\cite{ALICE-pp-2011,ALICE-pp-2012,ALICE-pp-2018,ALICE-pp-2020} collaborations as in the expression below,
\begin{equation} \label{Tsallis:non-ther}
\frac{1}{2\pi p_{\rm T}} \frac{d^2N}{dp_{\rm T} dy} = \frac{1}{2\pi} \frac{dN}{dy} \frac{(n-1)(n-2)}{n C(n C +m (n-2))} \left(1+\frac{m_{\rm T} -m}{n C} \right)^{-n},
\end{equation}
where $n,~C$ are the fitting parameters and $m$ is the rest mass of the particle. At mid-rapidity and zero chemical potential Eq.~\ref{Tsallis:non-ther} has same dependency on transverse momentum as in Eq.~\ref{Tsallis:rapidity} except the additional $m_{\rm T}$ dependency. The incorporation of $m_{\rm T}$ in Eq.~\ref{Tsallis:rapidity} has more consistent behavior of $q$ and $T$ whereas no clear pattern of $n$ and $C$ can be found~\cite{Cleymans-Worku-2012-EPJ}. In this present study, Tsallis distribution of thermodynamical consistent form given in Eq.~\ref{Tsallis:rapidity} has been used.
The conversion from rapidity to pseudo-rapidity phase-space is made using a Jacobian ($J(y,\eta)$) of the form,
\begin{equation} \label{Jacob}
\frac{dN}{dp_{\rm T} d\eta} = \sqrt{1-\frac{m^2}{m_{\rm T}^2\cosh^2(y)}} \frac{dN}{dy dp_{\rm T}}.
\end{equation}
At mid-rapidity  ($y \approx 0$), this can be simplified as 
\begin{equation} \label{Jacob-rapidity0}
\frac{dN}{dp_{\rm T} d\eta} = \frac{p_{\rm T}}{m_{\rm T}} \frac{dN}{dy dp_{\rm T}}. 
\end{equation}

The expression of the Tsallis distribution for all charged particle pseudo-rapidity distribution at mid-rapidity can be given as,
\begin{equation} \label{Tsallis:pseudo-rap}
\frac{d^2N_{\rm ch}}{dp_{\rm T} d\eta} = 2 p_{\rm T}^2 \frac{V}{(2\pi)^2} \sum_{i=1}^{3} g_i \left( 1+ ( q-1)\frac{m_{\rm T, i} -\mu}{T} \right)^{-\frac{q}{q-1}}. 
\end{equation}

Here, the sum runs for three kind of the particles $\pi^+, K^+$, and $p$ considering that those are the most abundant in productions. The factor $2$ in front of the right side takes into account of the associated anti-particles, $\pi^-, K^-$, and $\bar{p}$. The degeneracy factors ($g$) are unity for pions and kaons and $2$ for protons. At mid-rapidity, Eq.~\ref{Tsallis:rapidity} reduces to the form,
\begin{equation} \label{Tsallis:rapidity0}
\frac{1}{2\pi p_{\rm T}} \frac{d^2N}{dp_{\rm T} dy} = \frac{gV}{(2\pi)^3} m_{\rm T} \left( 1+ ( q-1)\frac{m_{\rm T} -\mu}{T} \right)^{-\frac{q}{q-1}}. 
\end{equation}

This equation (\ref{Tsallis:rapidity0}) has been  used to fit invariant momentum spectra of identified particles. 
The centrality dependent finite chemical potential $\mu$ has been
considered for the RHIC energy~\cite{STAR-BES1,Debadeepti-thesis-2019}, whereas at LHC energies $\mu$ has been approximated to zero. 
\section{Fits of the $p_{\rm T}$ spectra} \label{pT_spectrum_fitting}
\begin{figure*}[th!]
	\centering 
	\includegraphics[width=0.45\linewidth]{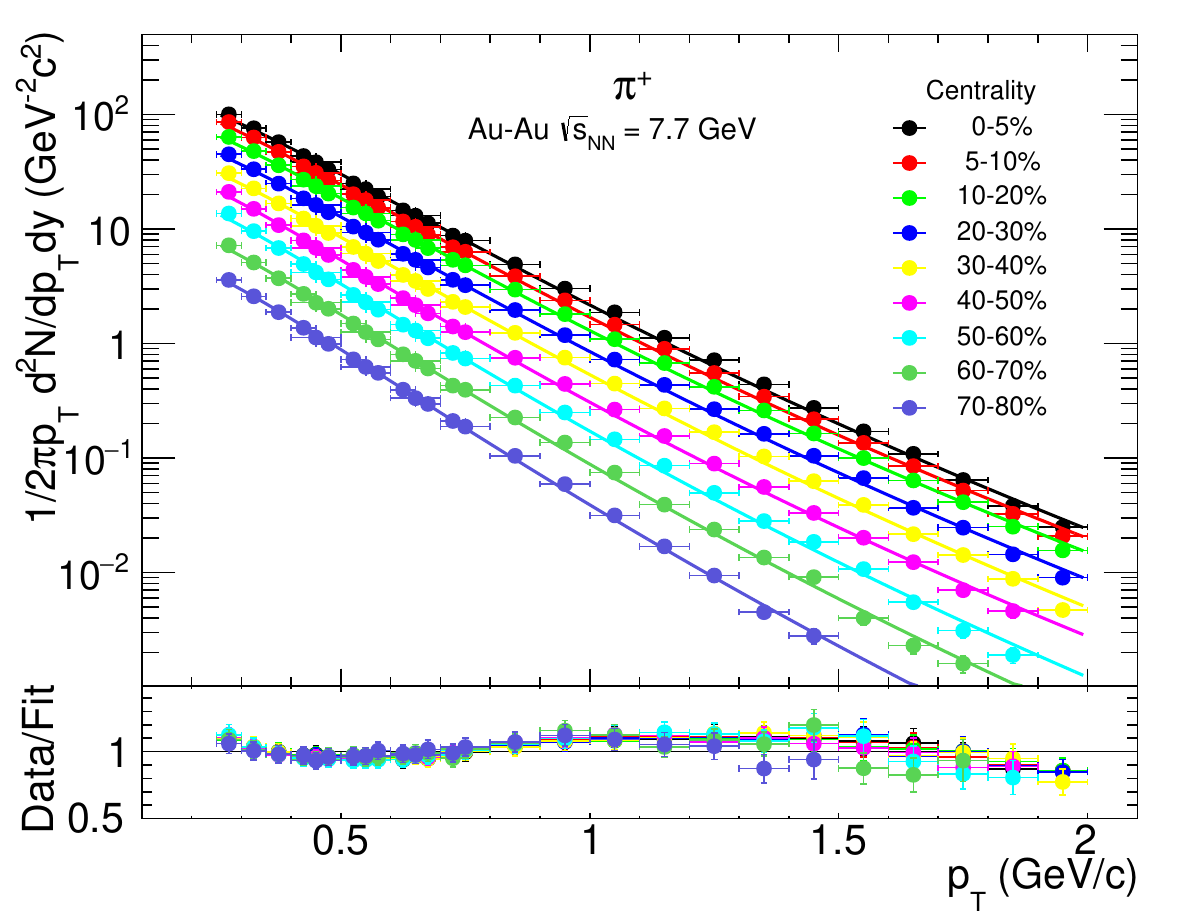}
	\includegraphics[width=0.45\linewidth]{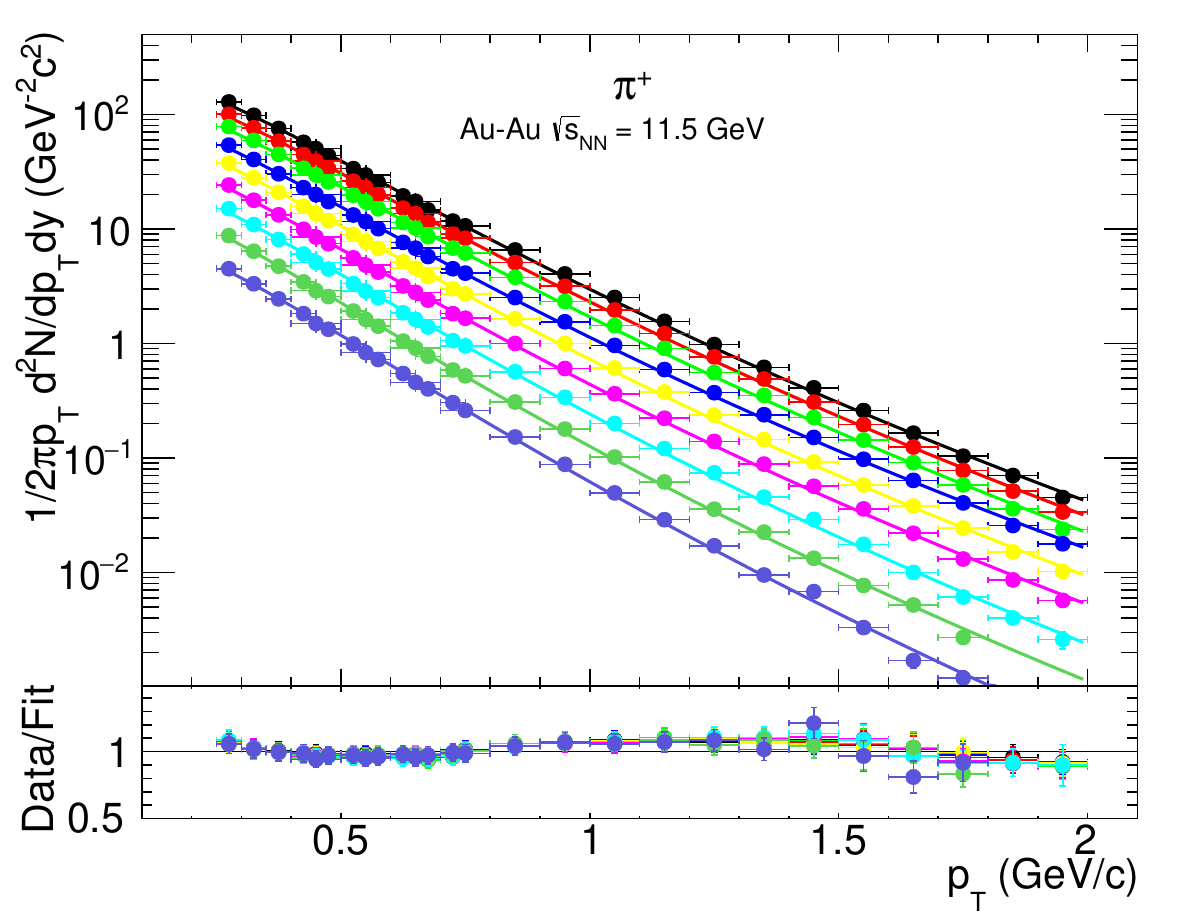}	
	\includegraphics[width=0.45\linewidth]{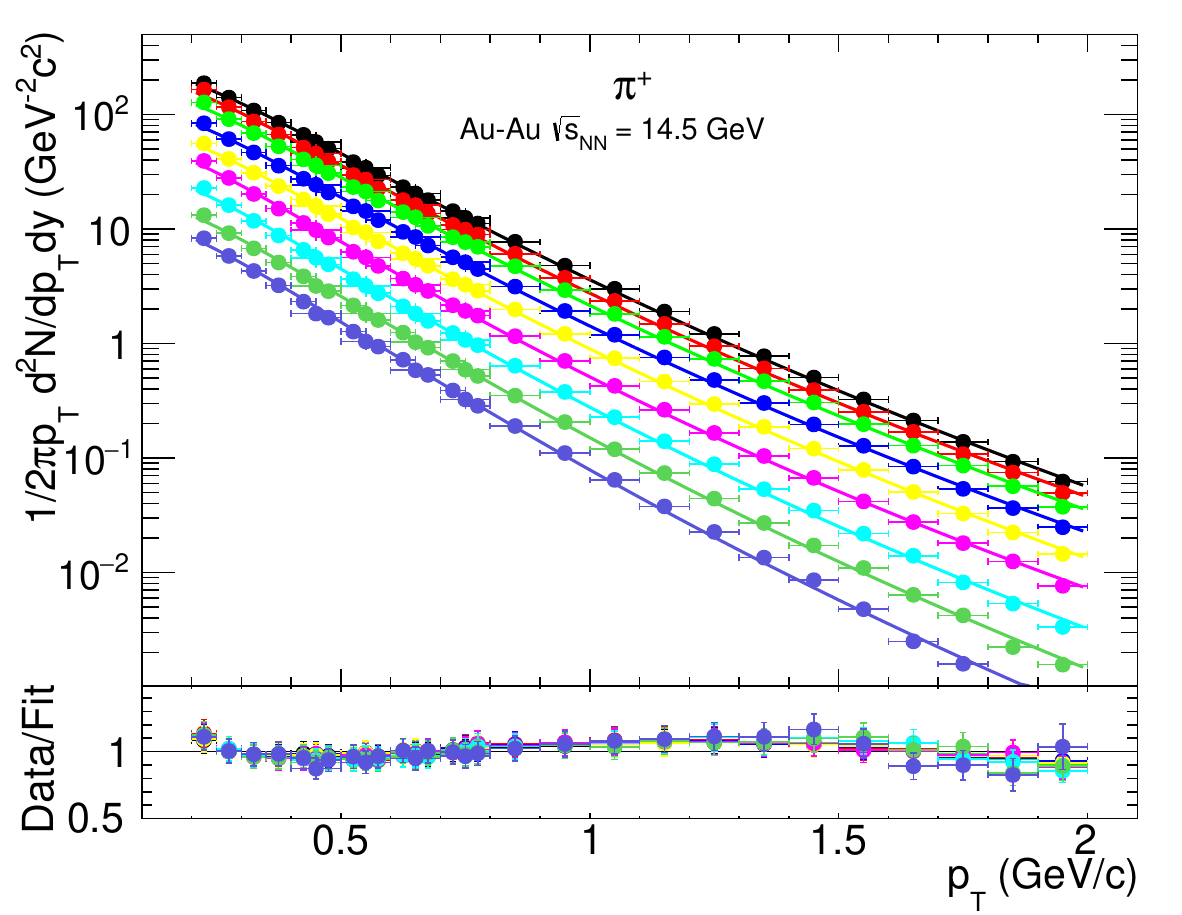}
	\includegraphics[width=0.45\linewidth]{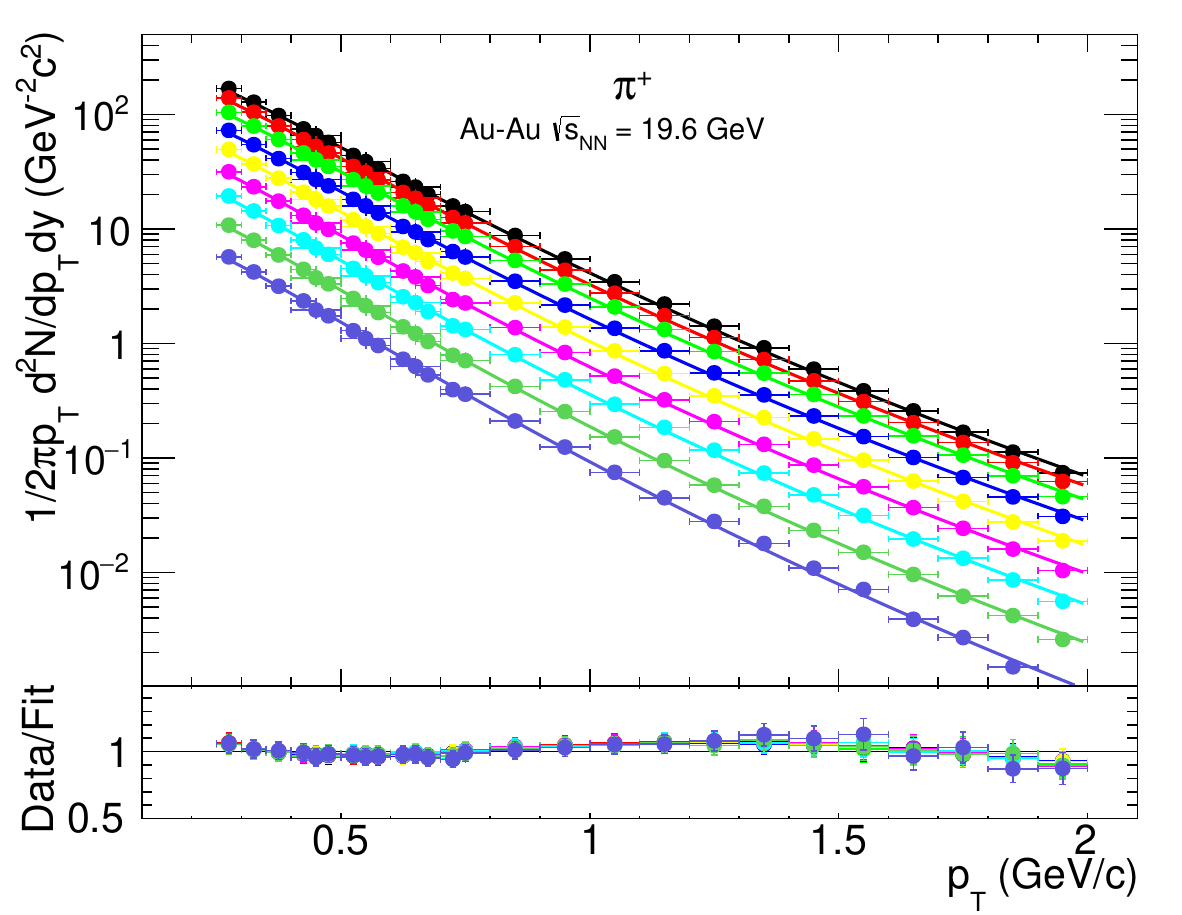}
	\includegraphics[width=0.45\linewidth]{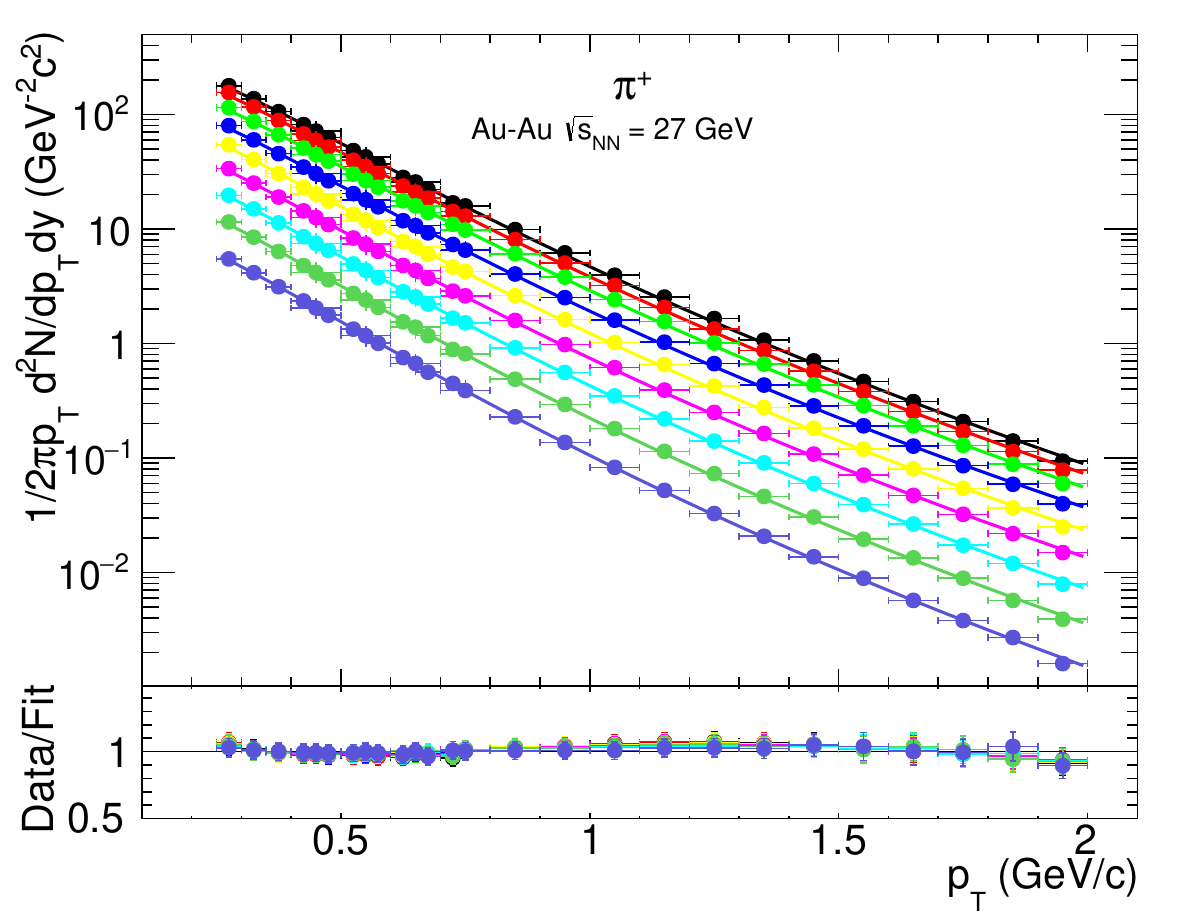}	
	\includegraphics[width=0.45\linewidth]{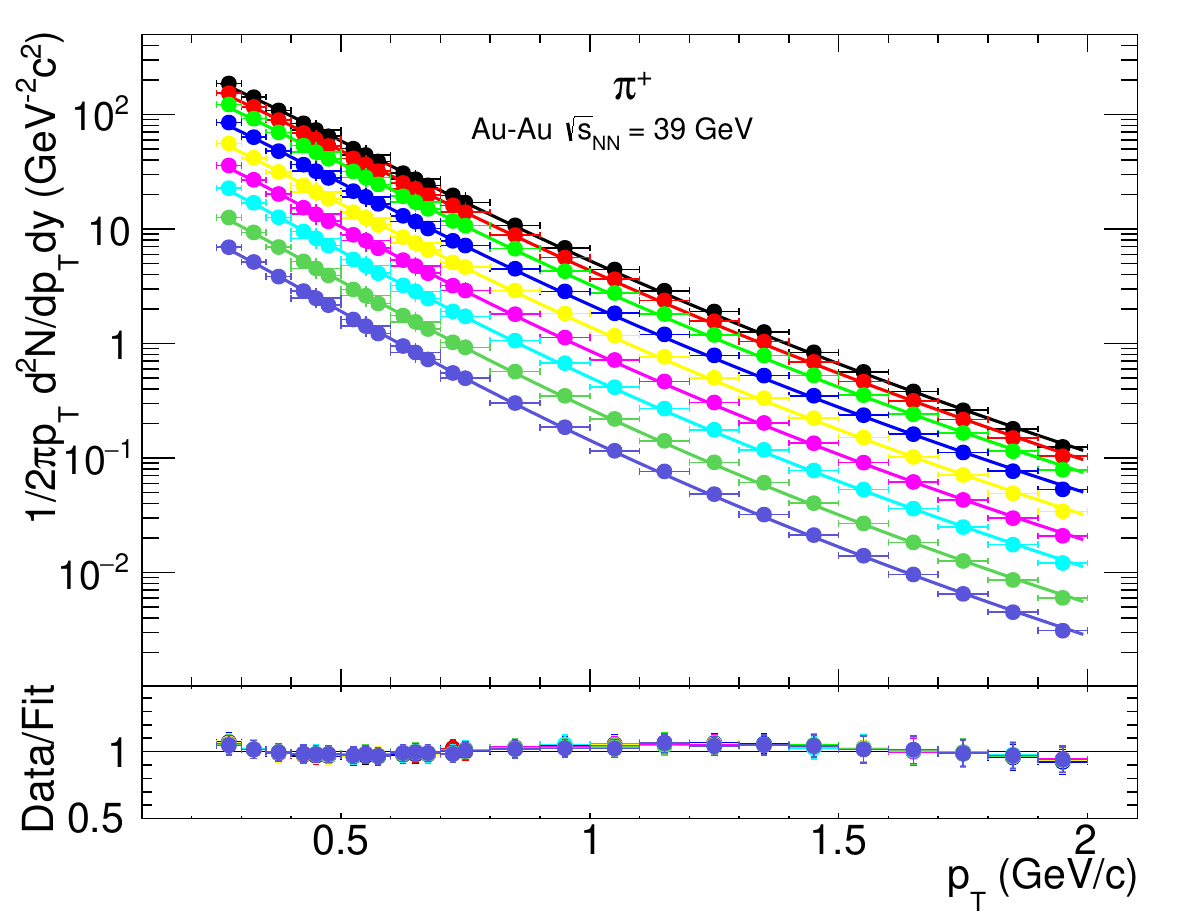}
\caption{Transverse momentum distributions of $\pi^+$ for different centralities at 
$\sqrt{s_{NN}}=$~7.7, 11.5, 14.5, 19.6, 27, 39~GeV, measured by the STAR experiment~\cite{STAR-BES1, STAR-14.5}. The solid lines represent fitting by the Tsallis distribution as given in Eq.~\ref{Tsallis:rapidity0}. The lower parts of the figures represent the ratios of the data to that of the fitting values.}
	\label{Spectrum:7-39GeV}
\end{figure*}
\begin{figure*}[th!]
	\centering 
	\includegraphics[width=0.45\linewidth]{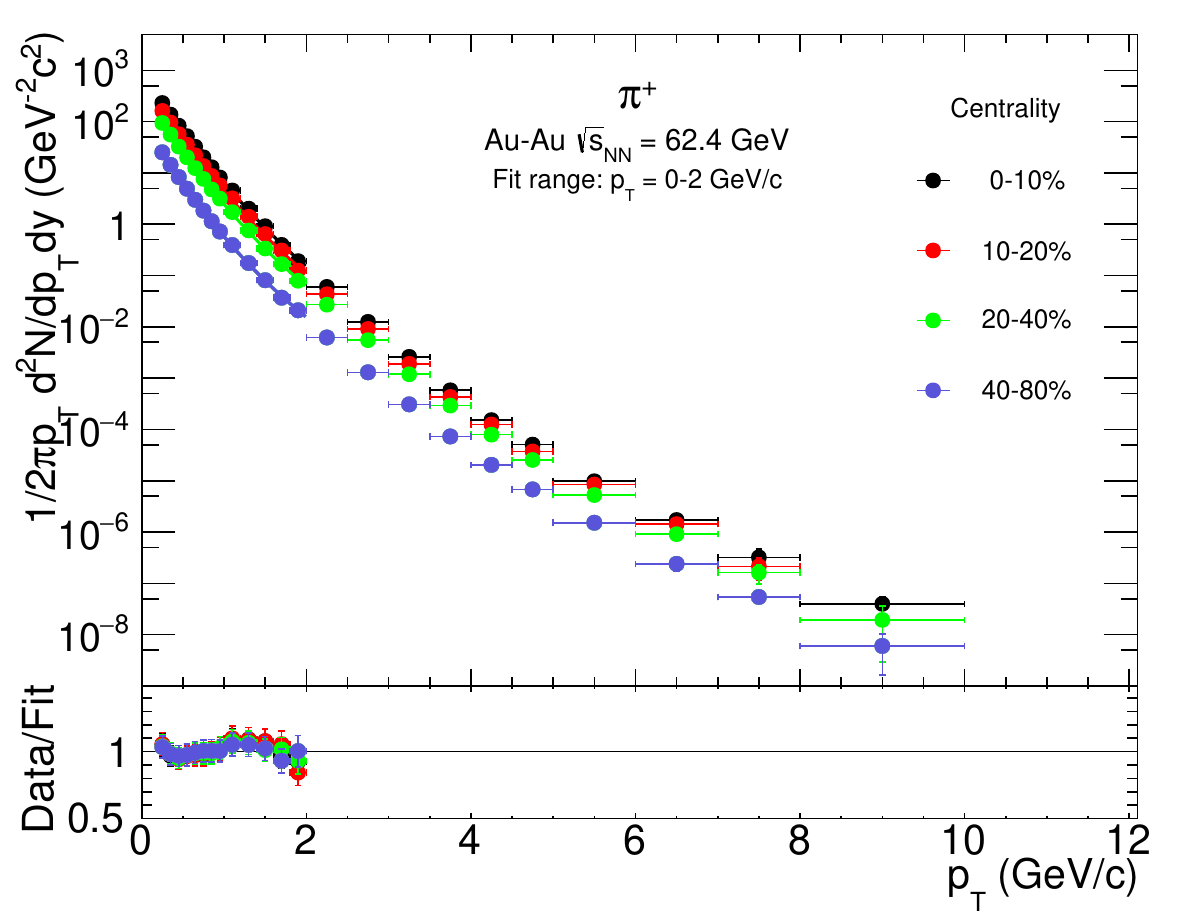}
	\includegraphics[width=0.45\linewidth]{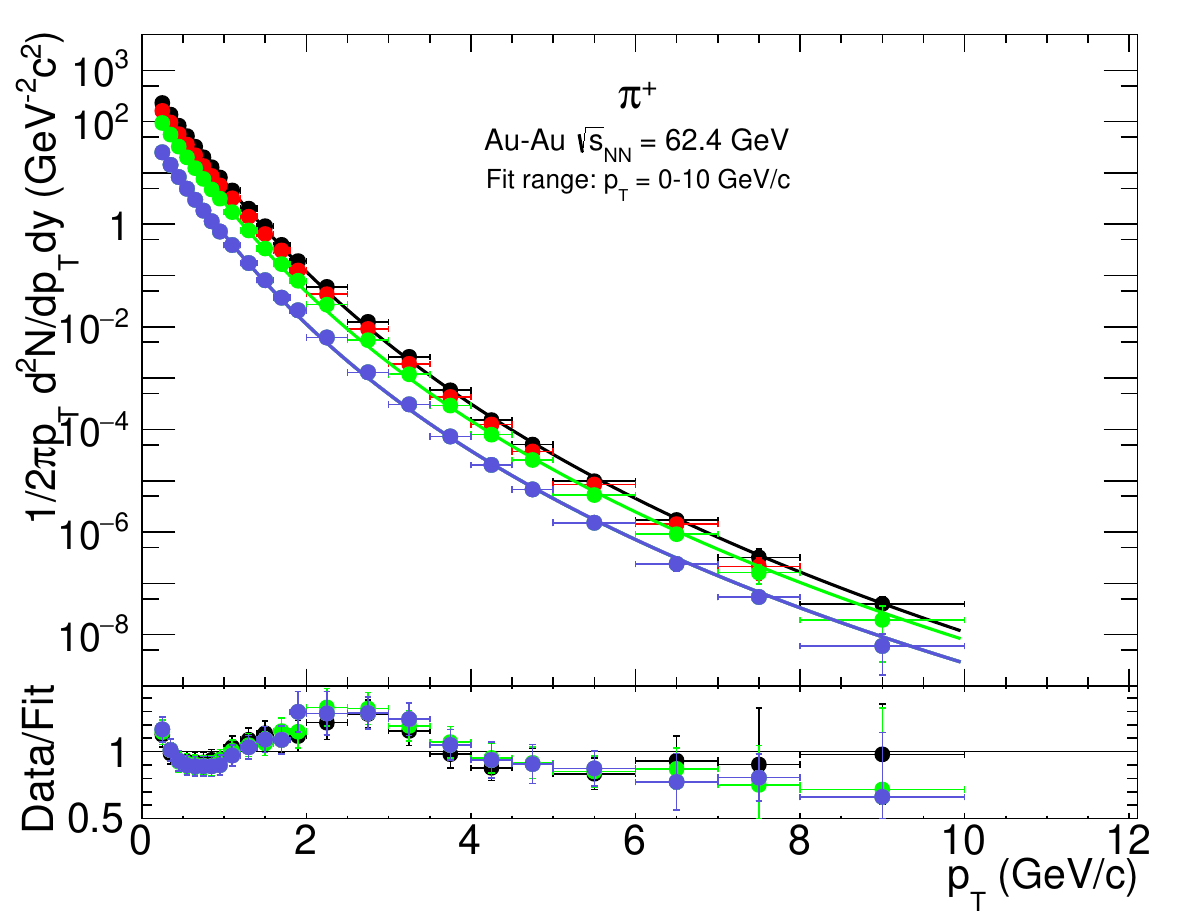}
	\includegraphics[width=0.45\linewidth]{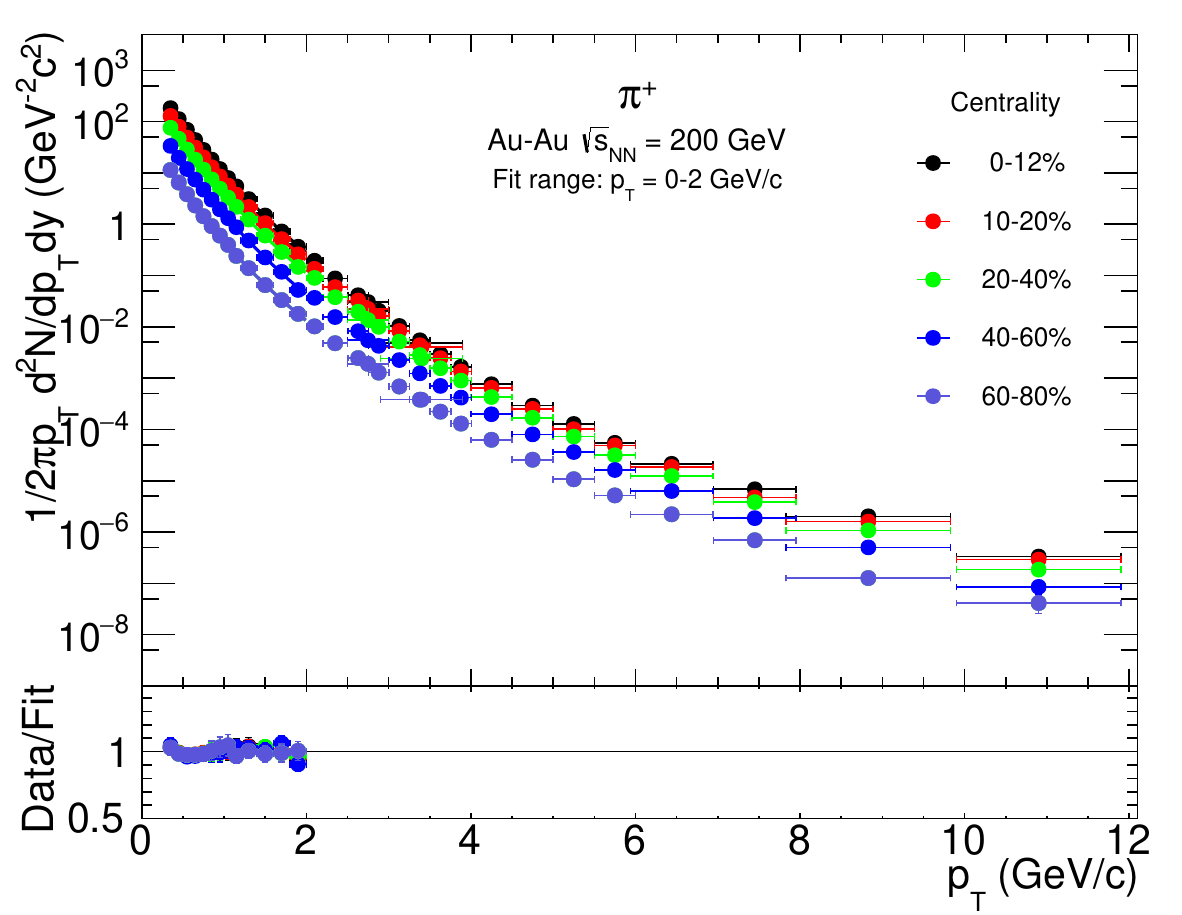}
	\includegraphics[width=0.45\linewidth]{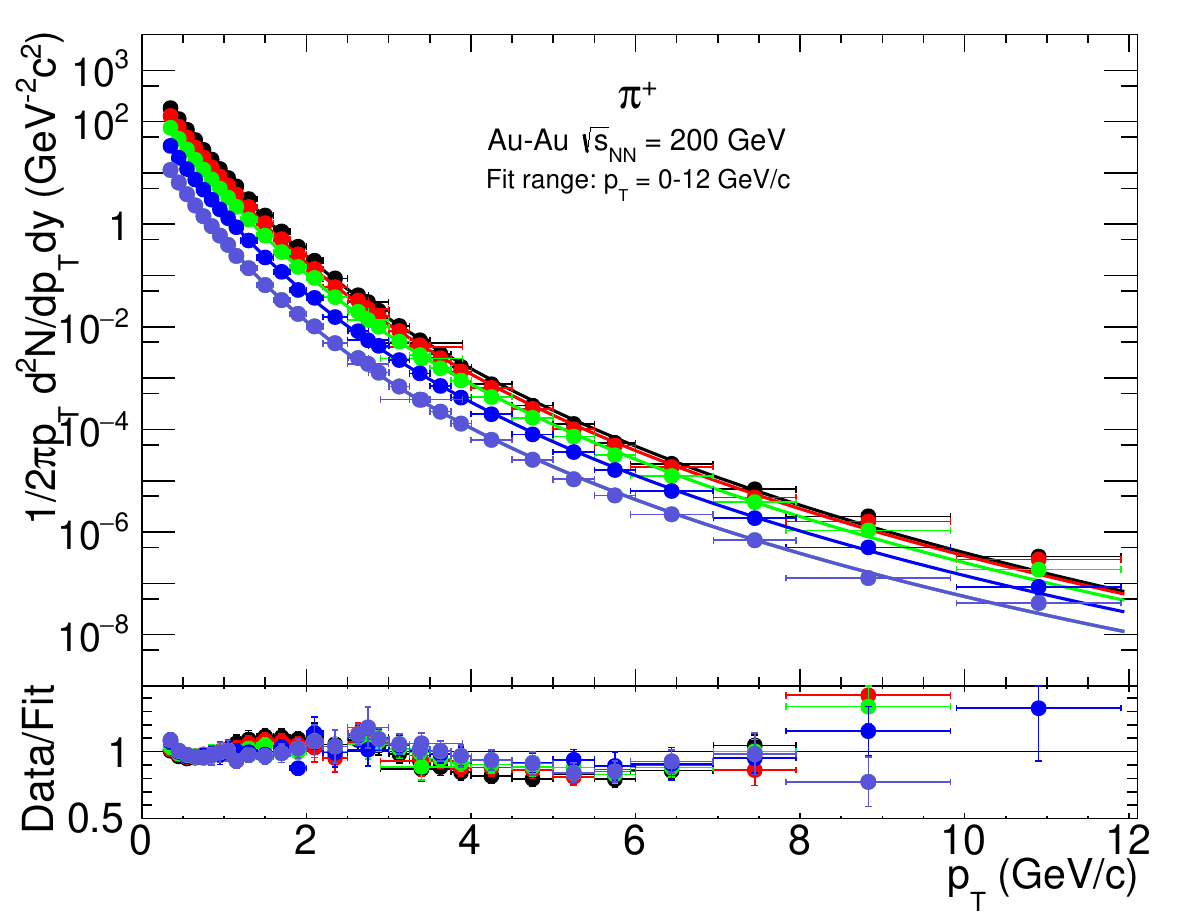}
	\caption{Transverse momentum distributions of $\pi^+$ for different centralities at $\sqrt{s_{NN}}=$~62.4 and 200~GeV measured by the STAR experiment~\cite{STAR-62.4,STAR-200}. The left and right panels correspond to fitting ranges of $p_{\rm T}$ up to 2~GeV/$c$ and up to the full $p_{\rm T}$ range, respectively. The solid lines represent fitting by the Tsallis distribution as given in Eq.~\ref{Tsallis:rapidity0}. The lower parts of the figures represent the the ratios of the data to that of the fitting values.} 
	\label{Spectrum:62-200GeV}
\end{figure*}
\begin{figure*}[th!]
	\centering 
	\includegraphics[width=0.32\linewidth]{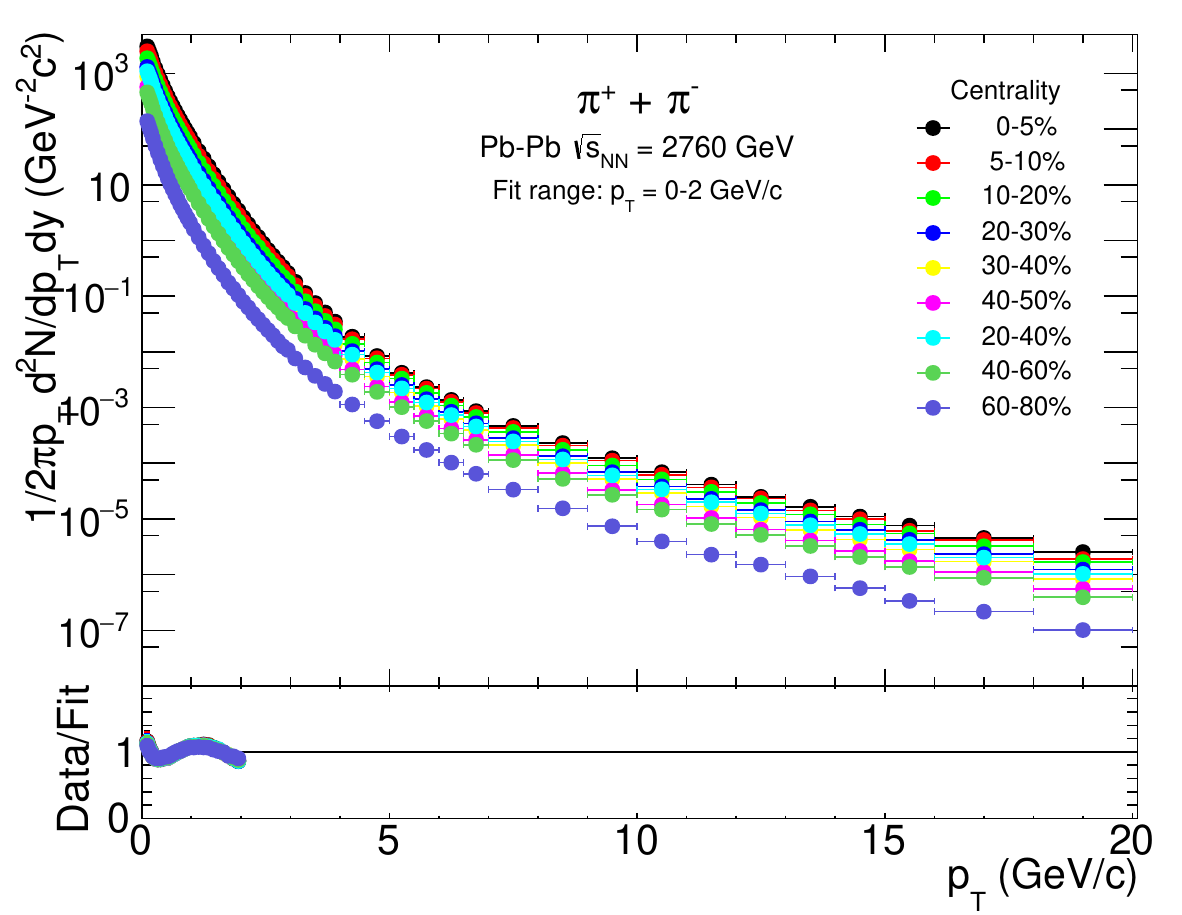}
	\includegraphics[width=0.32\linewidth]{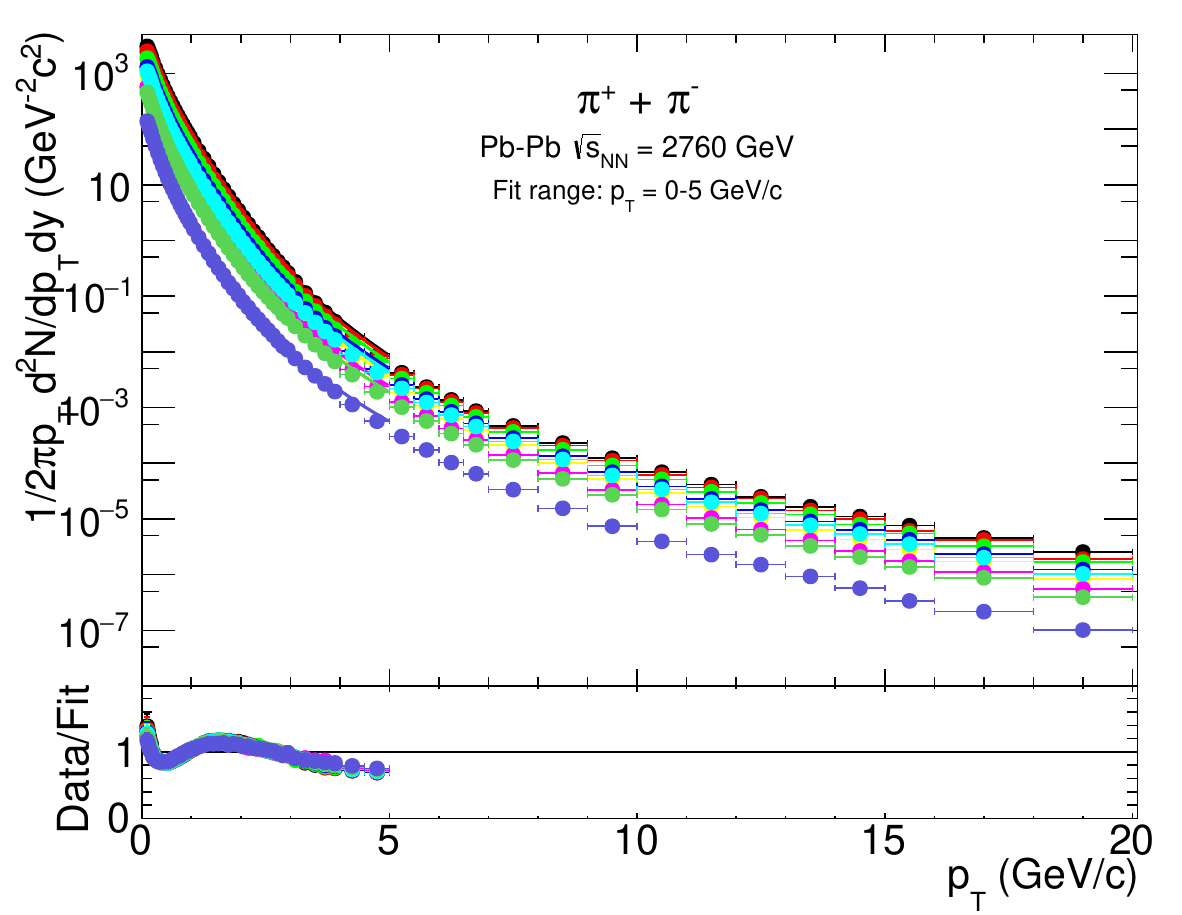}
	\includegraphics[width=0.32\linewidth]{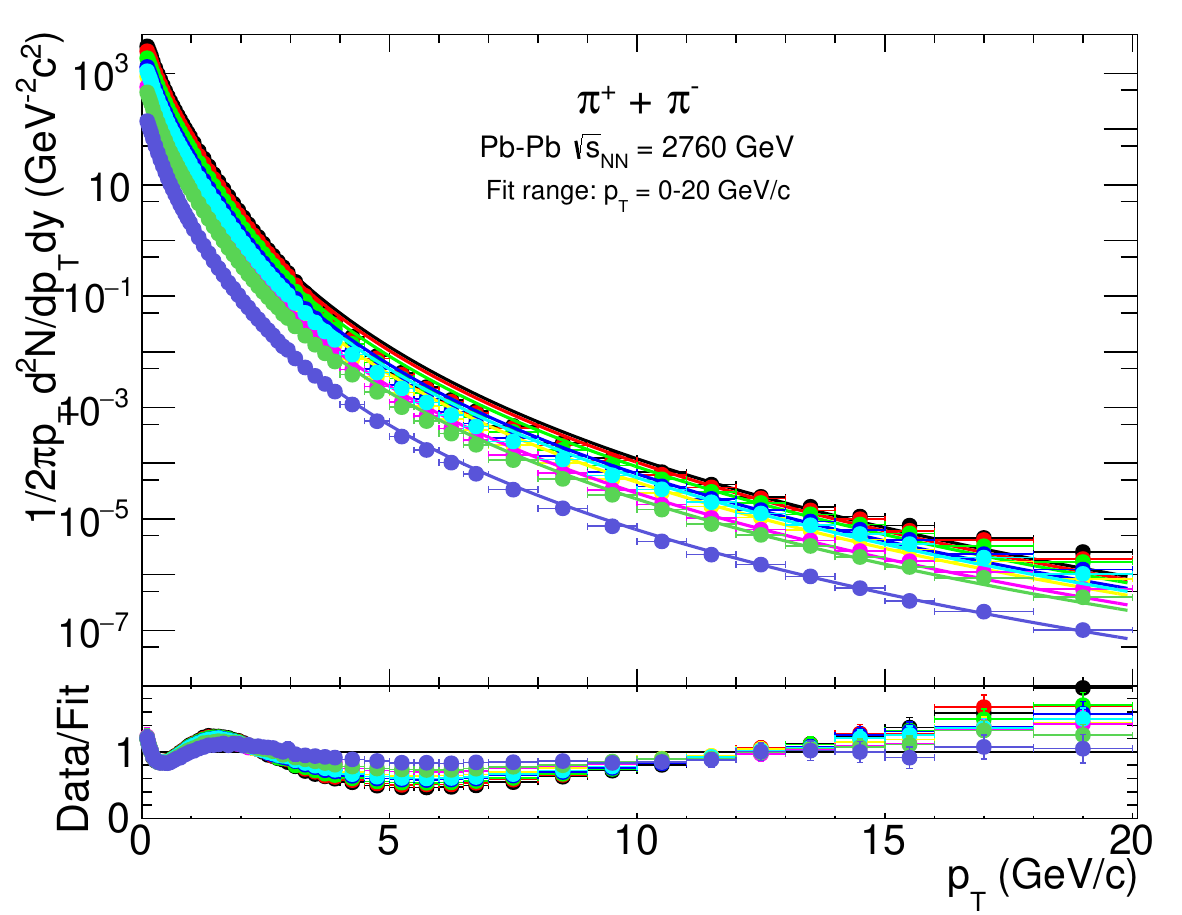}
	\caption{Transverse momentum distribution of $\pi^+ + \pi^-$ for different centralities at $\sqrt{s_{NN}}=$~2.76~TeV measured by the ALICE experiment~\cite{ALICE-2.76-2016}. The left, middle and right panels correspond to fitting ranges of  $p_{\rm T}$ up to 2~GeV/$c$, 5~GeV/$c$ and 20~GeV/$c$, respectively. The solid lines represent fitting by the Tsallis distribution as given in Eq.~\ref{Tsallis:rapidity0}. The lower parts of the figures represent the ratios of the data to that of the fitting values.}
	\label{Spectrum:2760GeV}
\end{figure*}

Detailed analysis of the transverse momentum spectra of produced
particles provides important information regarding the particle
production mechanisms. We have fitted the $p_{\rm T}$ spectra of
identified and all charged particles in heavy-ion collisions at RHIC
and LHC. At RHIC, the experimental data of Au-Au collisions from the
beam energy scan (BES-I) program are available for
$\sqrt{s_{NN}}=$~7.7, 11.5, 14.5 19.6, 27,
39~GeV~\cite{STAR-BES1,STAR-14.5}  as well as at 62.4 and at the top
energy of 200~GeV~\cite{STAR-62.4, STAR-200}.  Fine centrality binning
data from PHENIX collaboration at
$\sqrt{s_{NN}}=$~200~GeV~\cite{PHENIX-200} are also fitted. At the
LHC, data published by the ALICE collaboration for $p_{\rm T}$ spectra
of identified charged particles in Pb-Pb collisions at
$\sqrt{s_{NN}}=$~2.76~TeV~\cite{ALICE-2.76-2013,ALICE-2.76-2016} and
at 5.02~TeV~\cite{ALICE-5.02-2020} have been included in our study. In
addition, $p_{\rm T}$ spectra of all charged particles ($N_{\rm ch}$)
measured by ALICE at $\sqrt{s_{NN}}=$~2.76 and
5.02~TeV~\cite{ALICE-2.76-Nch,ALICE-5.02-Nch} are also used. In
Table~\ref{Summary: all_Energy}, we summarize the experimental
datasets used in the present study in terms of  collision energy, colliding system,
pseudo-rapidity ($\eta$) range, $p_{\rm T}$ range, and centrality
binning in percentage of the cross-section.
For the experimental data, the systematic and
statistical uncertainties are added in quadrature.

The fits to the transverse momentum spectra at all collision energies as given in Table~\ref{Summary: all_Energy} are performed by the expression of Tsallis distribution as in Eq.~\ref{Tsallis:rapidity0}. The transverse momentum spectra with Tsallis fitting of positively charged pions ($\pi^+$) at different centrality bins in Au-Au collisions at $\sqrt{s_{NN}}=$~7.7~GeV to 39~GeV measured by the STAR collaboration from the RHIC BES-I~\cite{STAR-BES1,STAR-14.5} program are shown in Fig.~\ref{Spectrum:7-39GeV}. The data are available for $p_{\rm T}$ range up to 2~GeV/$c$, and so the fits, represented by the solid lines, are also made up to this $p_{\rm T}$ range. The Tsallis distribution provides very good fits of the data at all centralities, which is evident from the ratio of the experimental data to the fit value as shown in the lower panel of the figure. The maximum deviations of the ratios from unity are within $\sim10-15\%$.  The fits of the $p_{\rm T}$ spectra of $\pi^-$ distributions are of similar quality to the $\pi^+$ distributions.

Fig.~\ref{Spectrum:62-200GeV} shows the $p_{\rm T}$ spectra along with the corresponding Tsallis fits for Au-Au collisions at $\sqrt{s_{NN}}=$~62.4 and 200~GeV from STAR~\cite{STAR-62.4,STAR-200} experiment with the available $p_{\rm T}$ ranges up to 10~GeV/$c$ and 12~GeV/$c$, respectively. The left panels of the figure show the fits of the $p_{\rm T}$ range up to 2~GeV/$c$ and the right panels provide the Tsallis fitting of the entire $p_{\rm T}$ spectrum. The ratios of the data to fit values are shown in the lower part of the figures. 
We observe that the quality of the Tsallis distribution fits shown in the left panels are quite good (with maximum deviations within $\sim10-15\%$) for fits up to low-$p_{\rm T}$ region of 2 GeV/c. Fits to higher $p_{\rm T}$ as shown in the right panels are not as good (the maximum deviations are larger, up to about 30\%). The fit parameters turn out to be dependent on the fit ranges in $p_{\rm T}$. The results of the fits are discussed in detail in the next section.

In Fig.~\ref{Spectrum:2760GeV}, we present the transverse momentum
spectra of all charged pions ($\pi^+ + \pi^-$) for different
centrality bins for Pb-Pb collisions at $\sqrt{s_{NN}}=$~2.76~TeV from
the ALICE~\cite{ALICE-2.76-2016} collaboration. The fits with the
Tsallis distribution function have been shown for three different
$p_{\rm T}$ ranges, where the left, middle, and the right plots in the
figure correspond to $p_{\rm T}$ ranges up to 2~GeV/$c$, 5~GeV/$c$ and
20~GeV/$c$, respectively. The lower parts of the plots give the ratio
of the data to the fitted values. These ratios show that the fitting
is in good agreement with the data for peripheral events, whereas
some deviations have been observed for central collisions and at large
$p_{\rm T}$. Furthermore, it is observed that the fit parameters vary
according to the fitting ranges in $p_{\rm T}$. 
In addition, we have fitted the low-$p_{\rm T}$ region of these spectra by fixing the parameter $q=1$
which gives the standard BG distribution, (Eq.~\ref{eq-BG}). We
observed that, fixing the parameter $q$ to unity does not
provide a reasonable fit even at low-$p_{\rm T}$ region $i.e.$ transverse momentum spectrum cannot be explained by exponential function only. Different studies present that at low-$p_{\rm T}$ region the spectra are modified by resonance productions~\cite{Aleksas resonance}, which is not considered here in the Tsallis fitting. 

\begin{figure}
	\centering
	\includegraphics[width=0.9\linewidth]{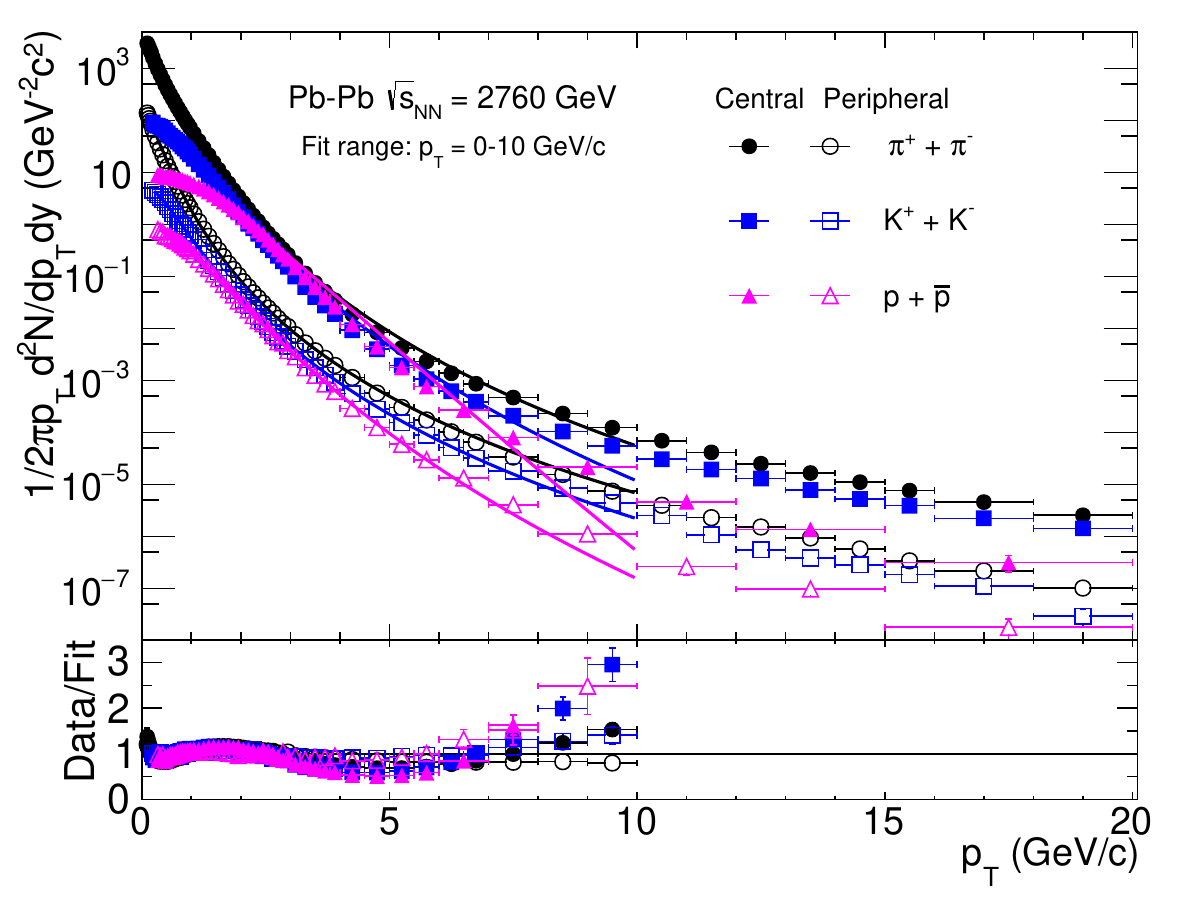}
	\caption{Transverse momentum distribution 
of $\pi^+ + \pi^-$, ${\rm K^+ + K^-}$ and ${\rm p +\bar{p} }$ for
most central (0--5\%, solid markers) and peripheral (70--80\%, open
markers) collisions at $\sqrt{s_{NN}}=$~2.76~TeV measured by the ALICE
experiment~\cite{ALICE-2.76-2016}. The solid lines represent the
fits up to $p_{\rm T}$ of 10~GeV/$c$ using the Tsallis
distribution as given in Eq.~\ref{Tsallis:rapidity0}. The lower panel
of the figure represent the ratios of the data to that of the fitting
values.}
\label{Fig:Pi-K-p_spectra_2.76}
\end{figure}

The transverse momentum spectra of identified charged pions, kaons and
protons at different collision energies and centralities 
are fitted using the Tsallis distribution.
The fitting of the $p_{\rm T}$ spectra of ${\rm K^+ +
  K^-}$ and ${\rm p +\bar{p} }$ along with $\pi^+ + \pi^-$ for Pb-Pb
collisions at $\sqrt{s_{NN}}=$~2.76~TeV~\cite{ALICE-2.76-2016} are
shown in Fig.~\ref{Fig:Pi-K-p_spectra_2.76} . The fits of the data
are performed for $p_{\rm T}$ ranges up to 10~GeV/$c$ for the most
central (0--5\%) and peripheral (70--80\%) collisions as presented by
the solid and open markers, respectively. The quality
of the fits is presented by the ratio of data to fit values at the
lower part of the figure. In general, we observe that the data for
peripheral collisions are fitted better compared to those of the
central collisions. Moreover, the fits are considerably better for
pions, then kaons and finally for protons.
The fits diverge gradually 
from the data with increasing particle masses and centrality at large
$p_{\rm T}$ ranges.

\section{Results of the Tsallis fit parameters}\label{results}
In this section, we present the results of the fitting of $p_{\rm T}$
distributions as a function of collision centrality, collision energy
as well as the $p_{\rm T}$ ranges of the fits. 
The results are presented for the Tsallis parameters, $q$, $T$, and the normalization volume, $V$. 
\subsection{Centrality dependence of Tsallis parameters}
\begin{figure}[th!]
	\centering 
	\includegraphics[width=0.9\linewidth]{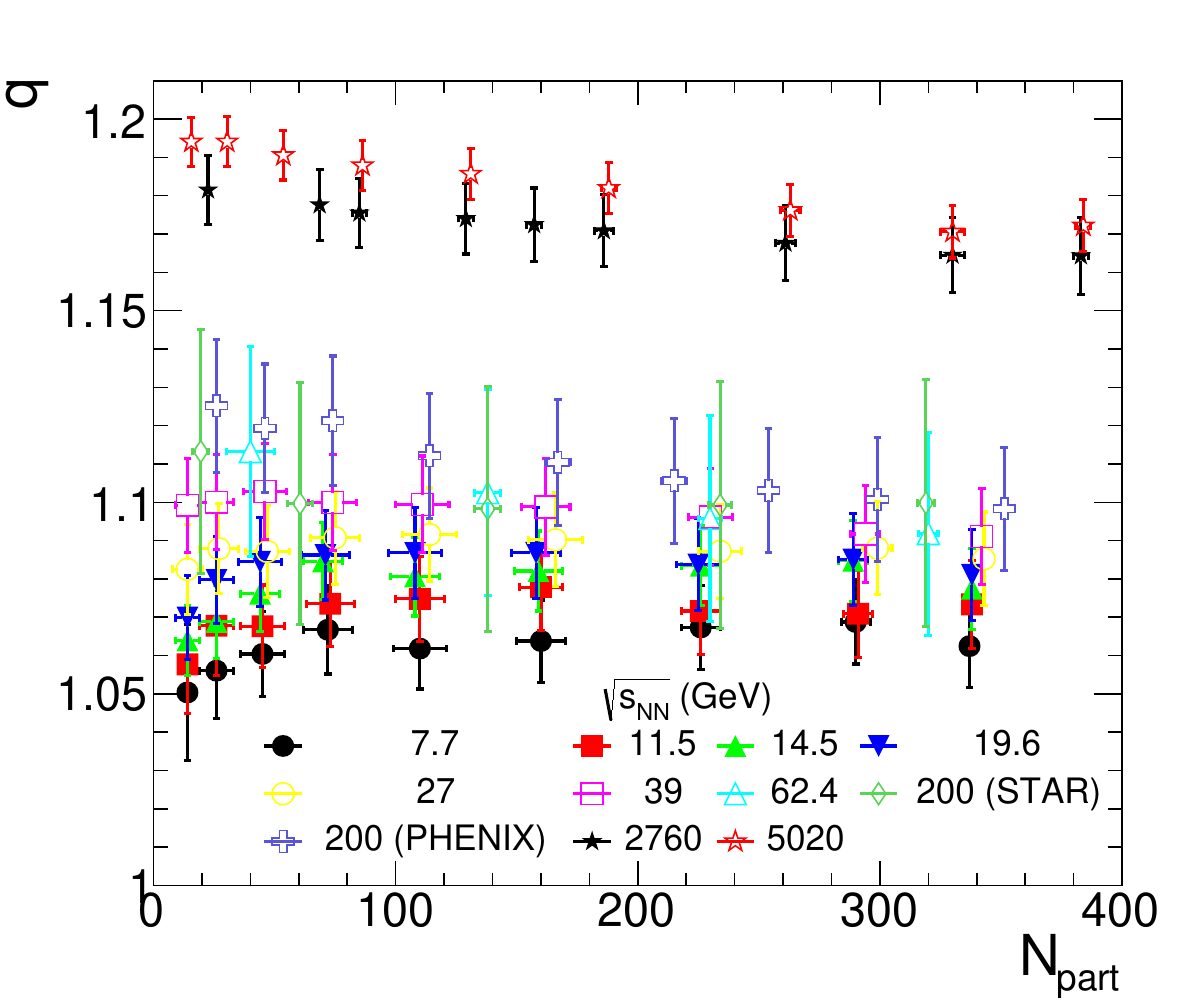}
	\includegraphics[width=0.9\linewidth]{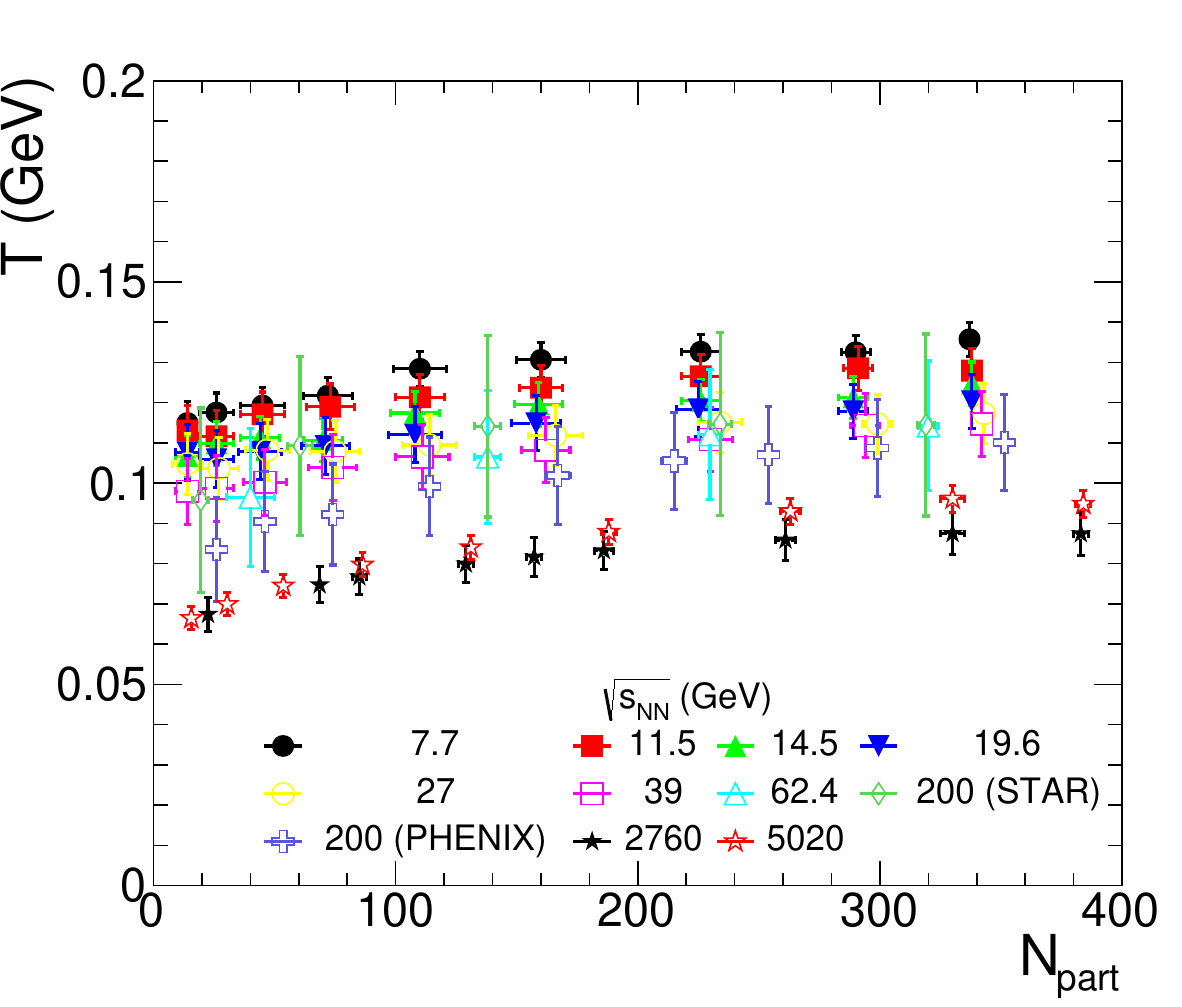}
	\includegraphics[width=0.9\linewidth]{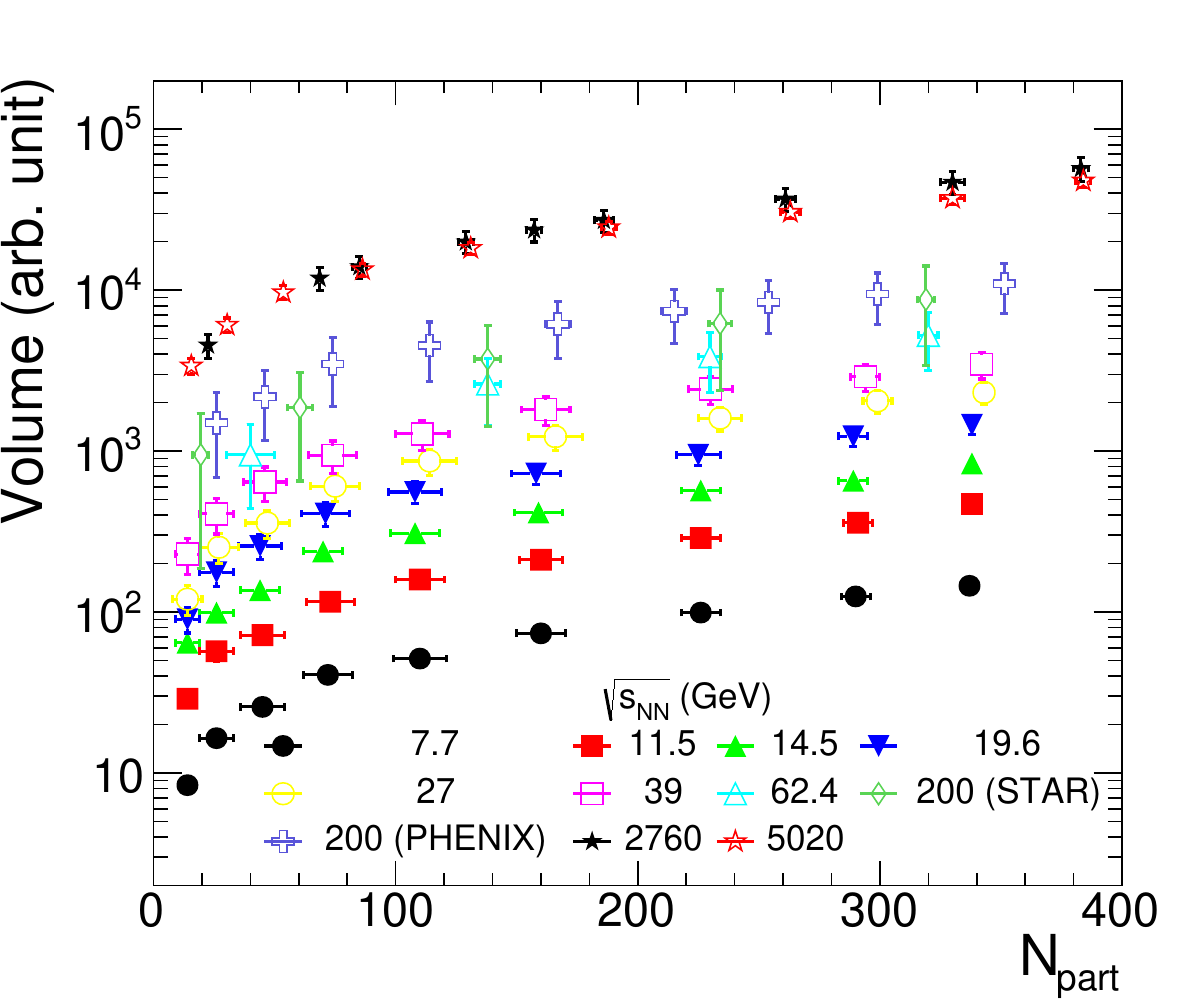}
	\caption{Tsallis parameters, $q$, $T$, and the normalization factor, $V$ as a function of number of participants, $N_{\rm part}$, obtained by fitting the $p_{\rm T}$ spectra up to 2~GeV/$c$ for $\pi^+$ at RHIC energies from STAR~\cite{STAR-BES1,STAR-14.5, STAR-62.4,STAR-200}, and PHENIX~\cite{PHENIX-200}, and all charged pions at LHC energies from ALICE~\cite{ALICE-2.76-2016,ALICE-5.02-2020}.
	}
	\label{Fig:q-Npart}
\end{figure}

\begin{figure}[htbp]
	\centering 
	\includegraphics[width=0.9\linewidth]{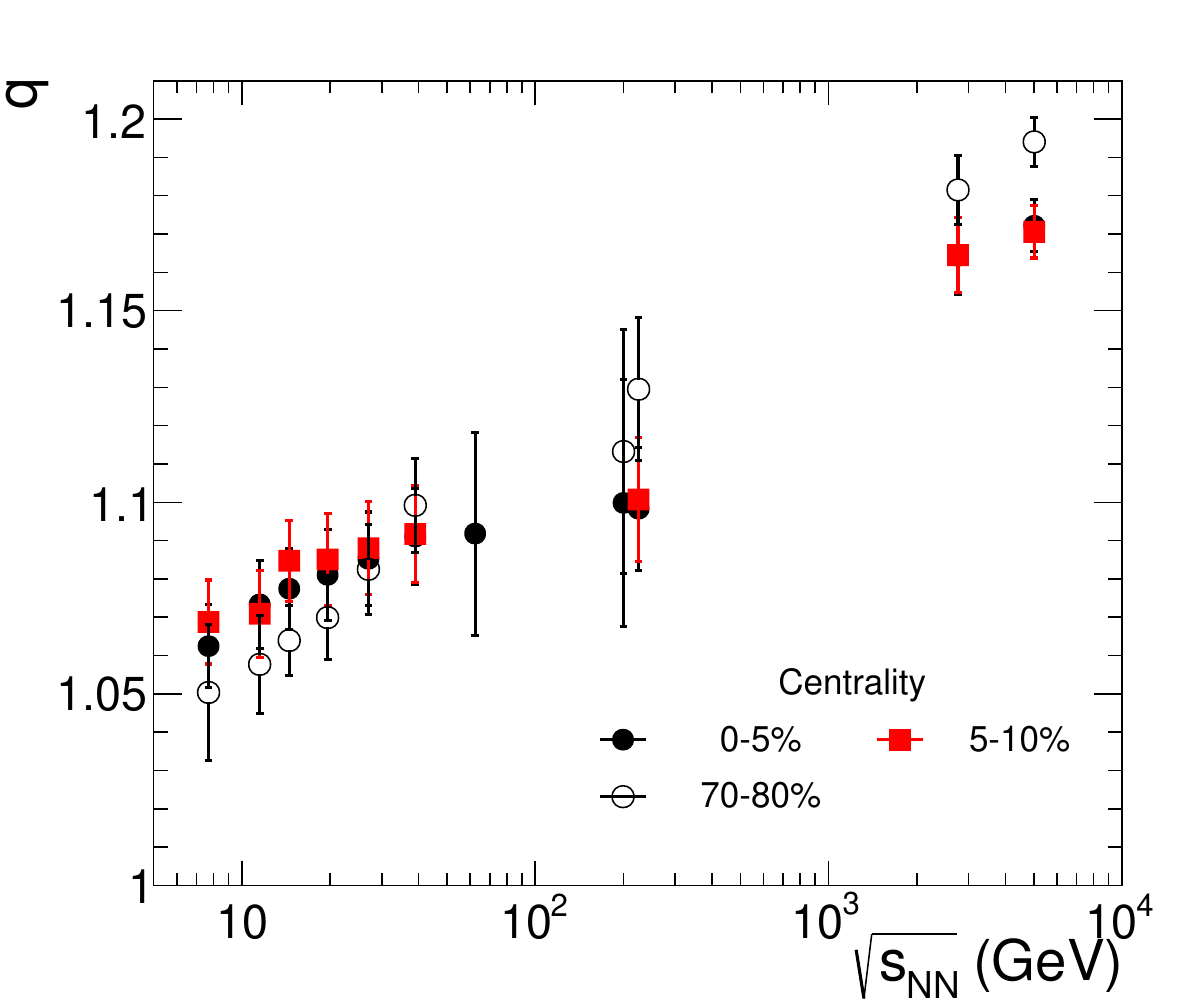}
	\includegraphics[width=0.9\linewidth]{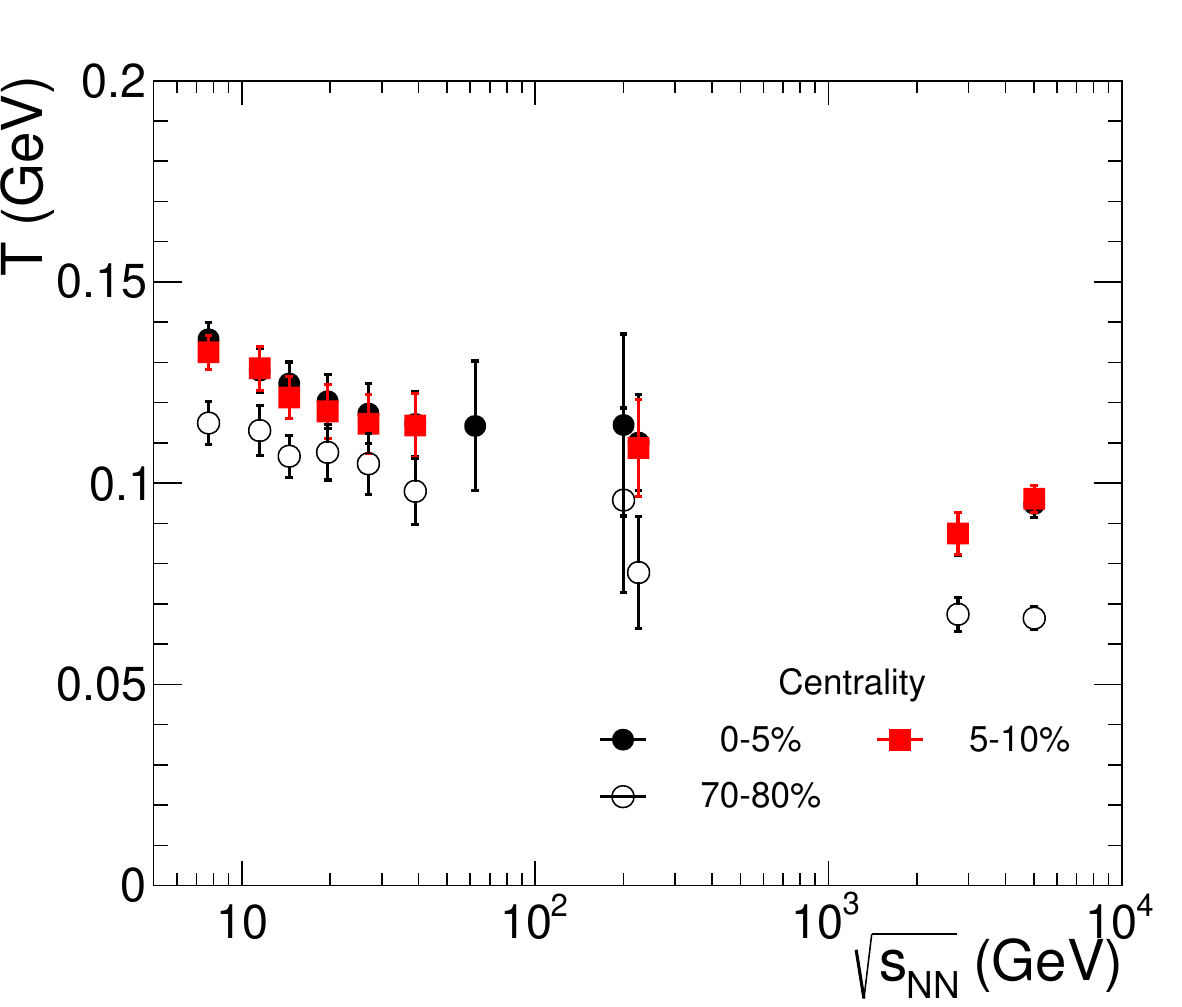}
	\includegraphics[width=0.9\linewidth]{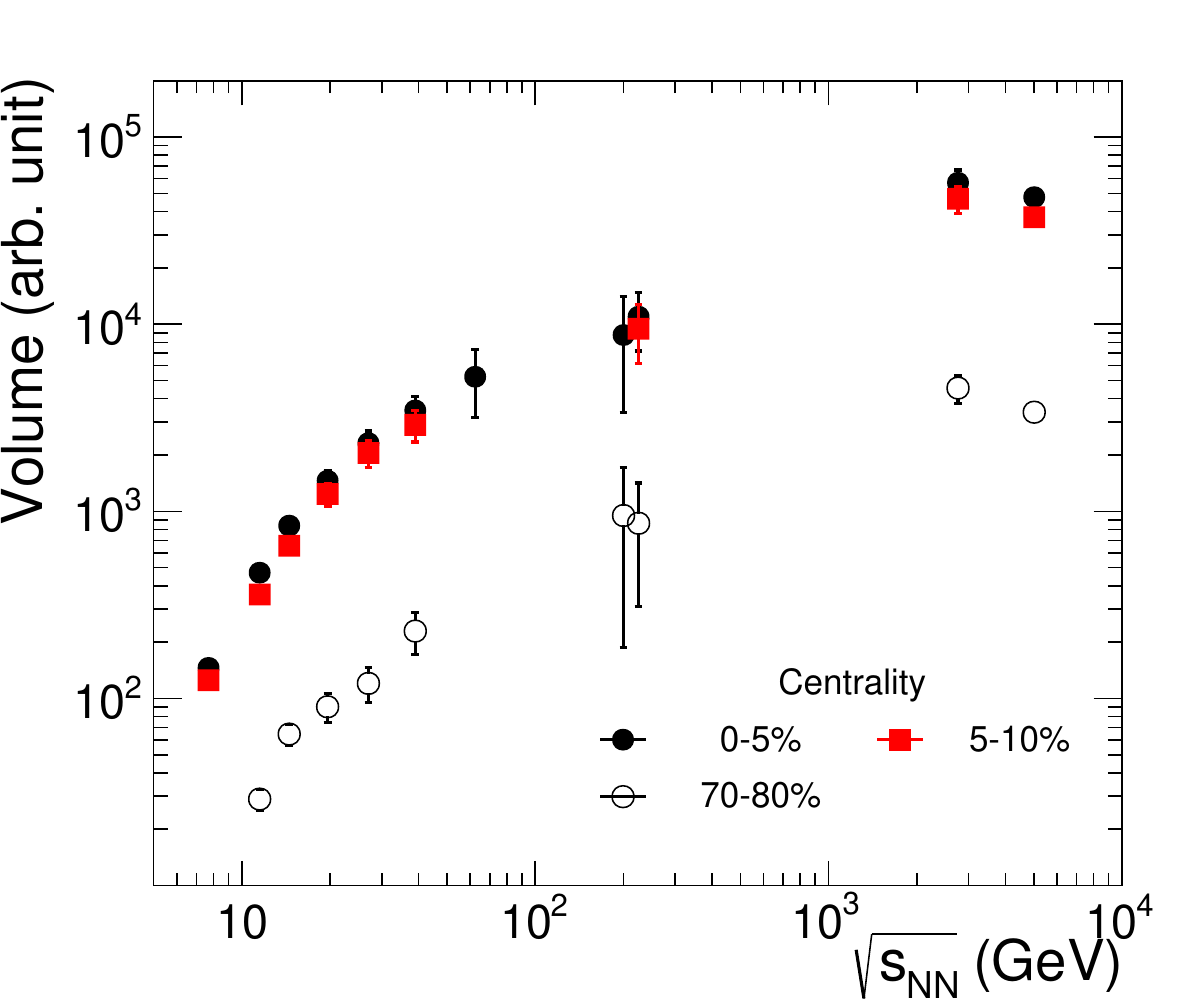}
	\caption{Tsallis parameters $q$, $T$, and the normalization factor, $V$ as a function of collision energy, $\sqrt{s_{NN}}$, for three different centralities obtained by fitting the $p_{\rm T}$ spectra up to 2~GeV/$c$ for $\pi^+$ at RHIC energies from STAR~\cite{STAR-BES1,STAR-14.5,STAR-62.4, STAR-200}, and PHENIX~\cite{PHENIX-200}, and all charged pions at LHC energies from ALICE~\cite{ALICE-2.76-2016,ALICE-5.02-2020}. PHENIX results are shifted to right for clarity. 
}
	\label{Fig:q-s}
\end{figure}

In the previous section, we have presented $p_{\rm T}$ distributions for different collision energies and collision centralities. The collision centralities have been shown in terms of the percentage of the cross-sections. For a given centrality window, the centrality can also be  expressed in terms of average number of participating nucleons ($N_{\rm part}$). As the centrality bins in the data as shown in Table~\ref{Summary: all_Energy} are different, it is useful to express the centrality in terms of $N_{\rm part}$. In Fig.~\ref{Fig:q-Npart}, we present the variation of $q$, $T$, and $V$ as a function of $N_{\rm part}$ for positively charged pions at RHIC energies, and for all charged pions at LHC energies. The parameters obtained from $\pi^+$ and $\pi-$ spectra of ALICE data~\cite{ALICE-2.76-2013} are similar to the results for all charged pions.  All the fits are made for maximum values of $p_{\rm T}$ up to 2~GeV/$c$. The following observations are made regarding the centrality and collision energy dependence of these parameters:
\begin{itemize}
\item {The Tsallis parameter, $q$, systematically increases from low
    to high collision energies. For each collision energy,  the variation of $q$ as a function of
    $N_{\rm part}$ has been inferred by fitting the dependence with a
    linear function. We observe that within the given uncertainly, $q$
    is independent centrality at RHIC energies. However, for
    LHC energies $q$ decreases from peripheral to central collisions.}
\item {The Tsallis temperature, $T$, decreases from low to high
    collision energies. For all collision energies, the values of $T$ increase from peripheral to central collisions.}
\item {The normalization parameter, $V$, increases from low to high collision energies. For all collision energies, $V$ increases monotonically from peripheral to central collisions.}
\end{itemize}
\subsection{Collision energy dependence of Tsallis parameters}
The variations of the Tsallis parameters, $q$, $T$, and the normalization parameter, $V$, are scanned over the collision energy $\sqrt{s_{NN}}$ from RHIC to LHC energies. The results of the fits to $p_{\rm T}$ spectra up to a value of 2~GeV/$c$ are presented in Fig.~\ref{Fig:q-s}. At RHIC energies, the fits are performed for $\pi^+$, and at the LHC energies, the fits are presented for $\pi^+ + \pi^-$. For clarity, results are presented only for central (0--5\%, 5--10\%) and peripheral (70--80\%) collisions. The following observations are made regarding the collision energy dependence of these parameters:
\begin{itemize}
\item{The parameter $q$ as a function of collision energy shows that for all centralities $q$ increases with the increase of collision energy. A closer look at the centrality dependence shows that at RHIC energy $q$ remains unchanged with the centrality within the uncertainty, whereas, in higher energies, $q$ decreases from peripheral to central collisions. }
\item{The parameter $T$ decreases with the increase of the collision energy. Another important observation is that $T$ increases from peripheral to central collisions at all collision energies.}
\item{The variation of $V$ with collision energy shows that, $V$ increases consistently with increasing $\sqrt{s_{NN}}$. This is  reasonable as the system size and hence particle productions increase with $\sqrt{s_{NN}}$ and centrality.}
\end{itemize}

\begin{figure}[!th]
	\centering
	\includegraphics[width=0.9\linewidth]{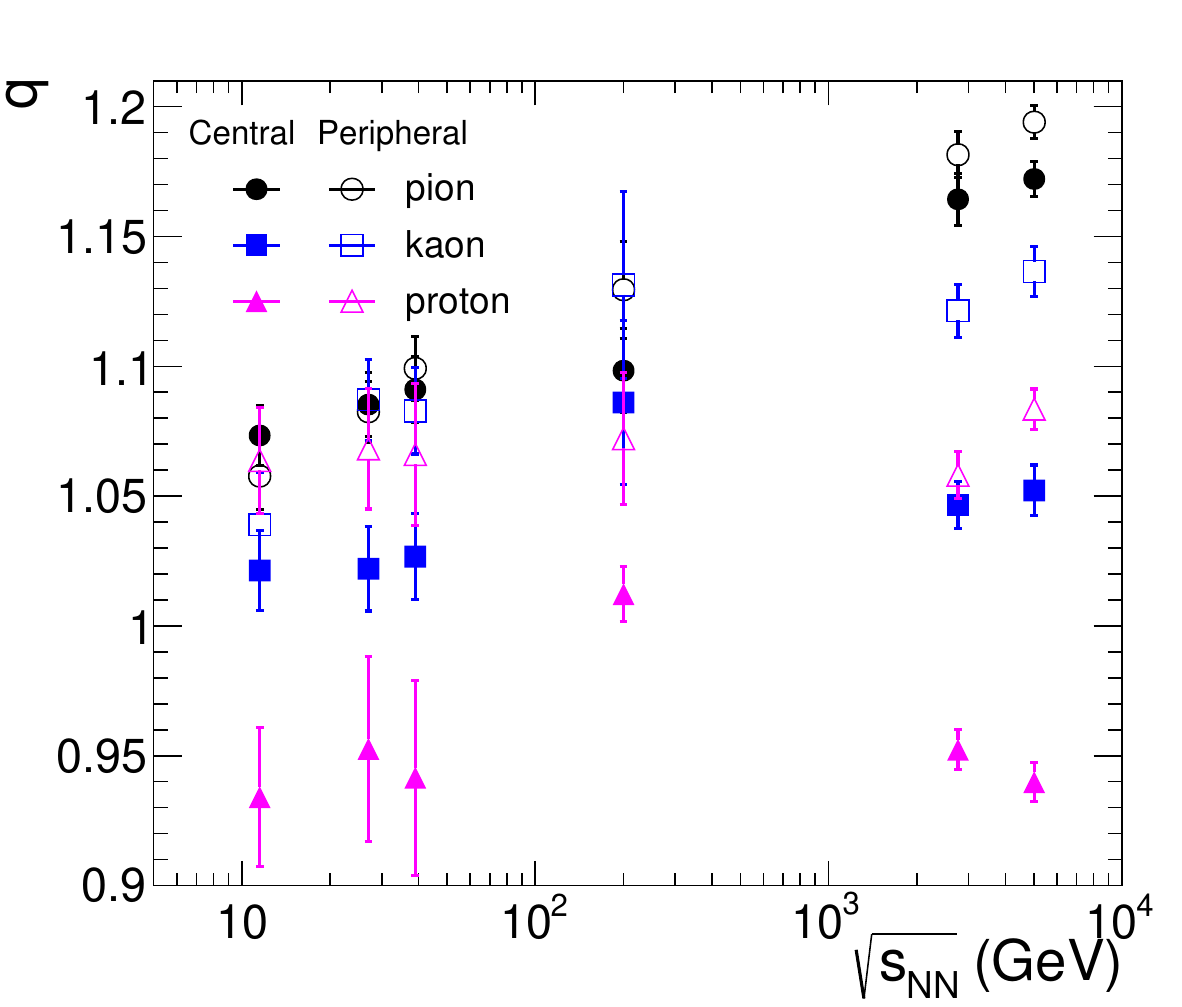}
	\includegraphics[width=0.9\linewidth]{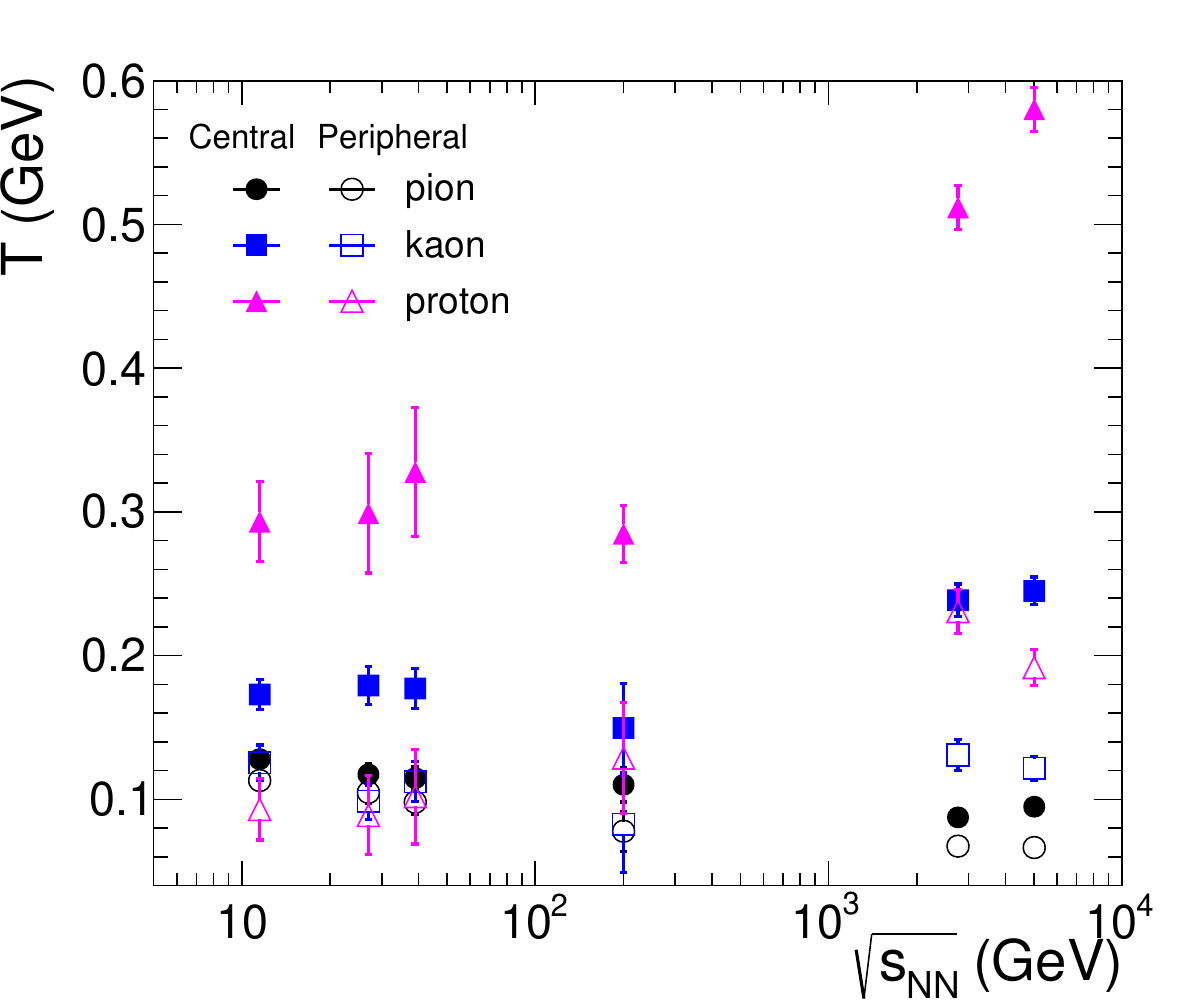}
	\caption{Tsallis parameters, $q$ and $T$ as a function of
          collision energy, $\sqrt{s_{NN}}$, for $\pi^+ (\pi^+ +
          \pi^-), \rm{K^+ (K^+ + K^-)}$ and p (p + $\bar{\rm p}$) for
          fits up to $p_{\rm T}$ range of 2~GeV/$c$, 3~GeV/$c$
          and 5~GeV/$c$, from STAR~\cite{STAR-BES1},
          PHENIX~\cite{PHENIX-200}, and
          ALICE~\cite{ALICE-2.76-2016,ALICE-5.02-2020} data. The
          open and solid markers correspond to peripheral and central collisions, respectively.}
	\label{Fig:q_T_PID}
\end{figure}

The usefulness of the Tsallis parameters can be understood by fitting the
$p_{\rm T}$ spectra of kaons and protons along with those of
pions. This has been done for RHIC energies 
from the STAR BES-I~\cite{STAR-BES1} and PHENIX~\cite{PHENIX-200}
datasets and at LHC for the
ALICE~\cite{ALICE-2.76-2016, ALICE-5.02-2020} data. The variations of
$q$ and $T$ as a function of $\sqrt{s_{NN}}$ are shown in
Fig.~\ref{Fig:q_T_PID} for all the three identified particles. The
parameters presented in the figure correspond to the 
$p_{\rm  T}$ range up to 2~GeV/$c$,  3~GeV/$c$ and 5~GeV/$c$ for
pions, kaons, and protons, respectively. The results of the peripheral and central collisions are labeled as open and solid markers, respectively. It has been
observed from the figure that for pions and kaons 
$q$ has an increasing trend with
$\sqrt{s_{NN}}$ for both central and peripheral collisions. 
But for protons, the trend is very different, and at the LHC energies $q$ values
are found lower than unity within the fitting range of the proton
spectra, which is against the basic assumption of the distribution function. This could be because of the radial flow which has a larger effect on protons that is not considered in the fitting.
The temperature $T$ shows a distinct difference for central and peripheral collisions for all the
three particle species. For peripheral collisions, $T$ values are similar for
all particles. But for 
central collisions, $T$ shows a strong mass dependency. For pions, $T$
decreases with increase of $\sqrt{s_{NN}}$, whereas $T$ increases for
kaons and protons, with unusually high value for protons.

\subsection{Behavior of Tsallis parameters for different $p_{\rm T}$ ranges}
\begin{figure}[th!]
	\centering 
	\includegraphics[width=0.9\linewidth]{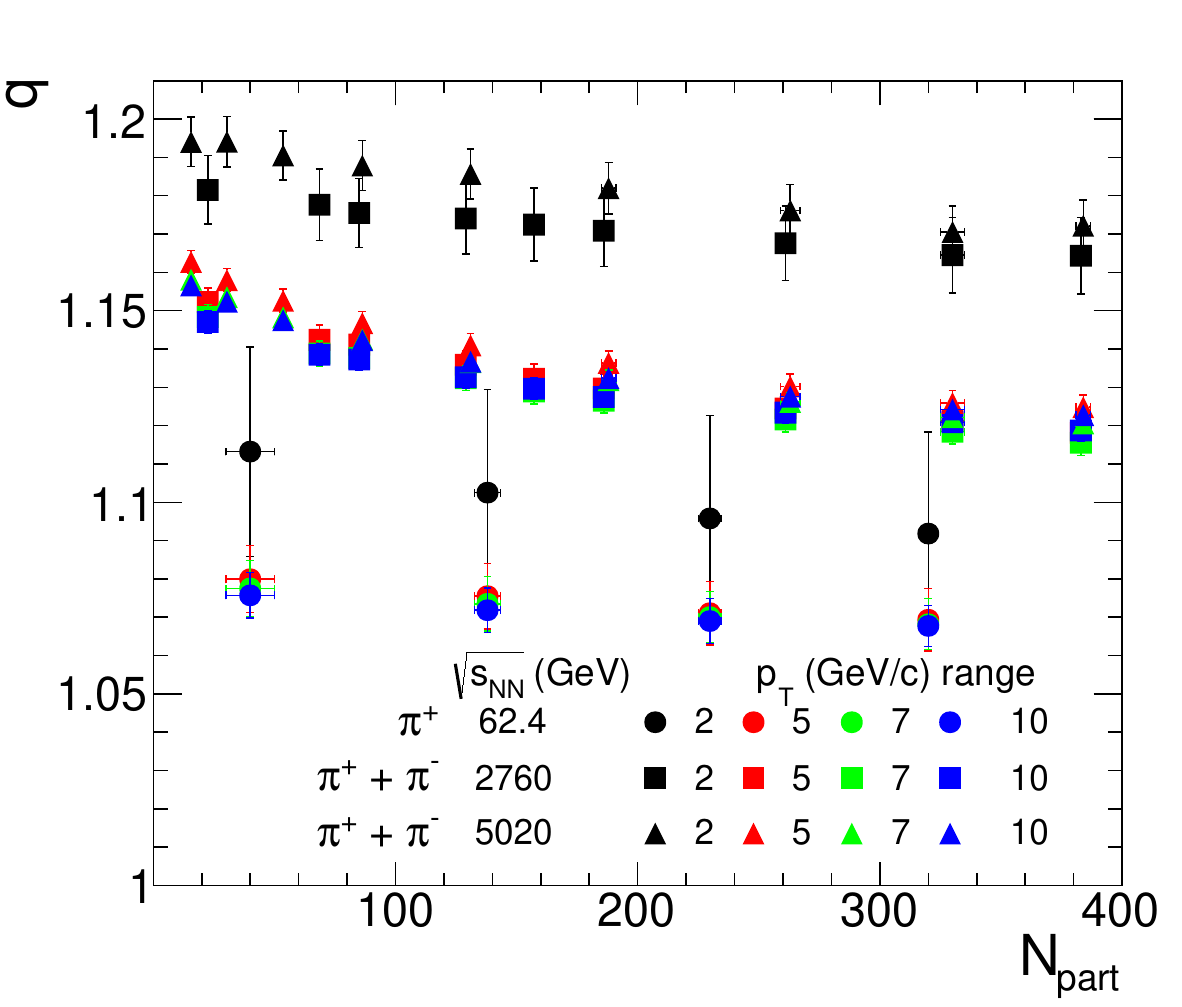}
	\includegraphics[width=0.9\linewidth]{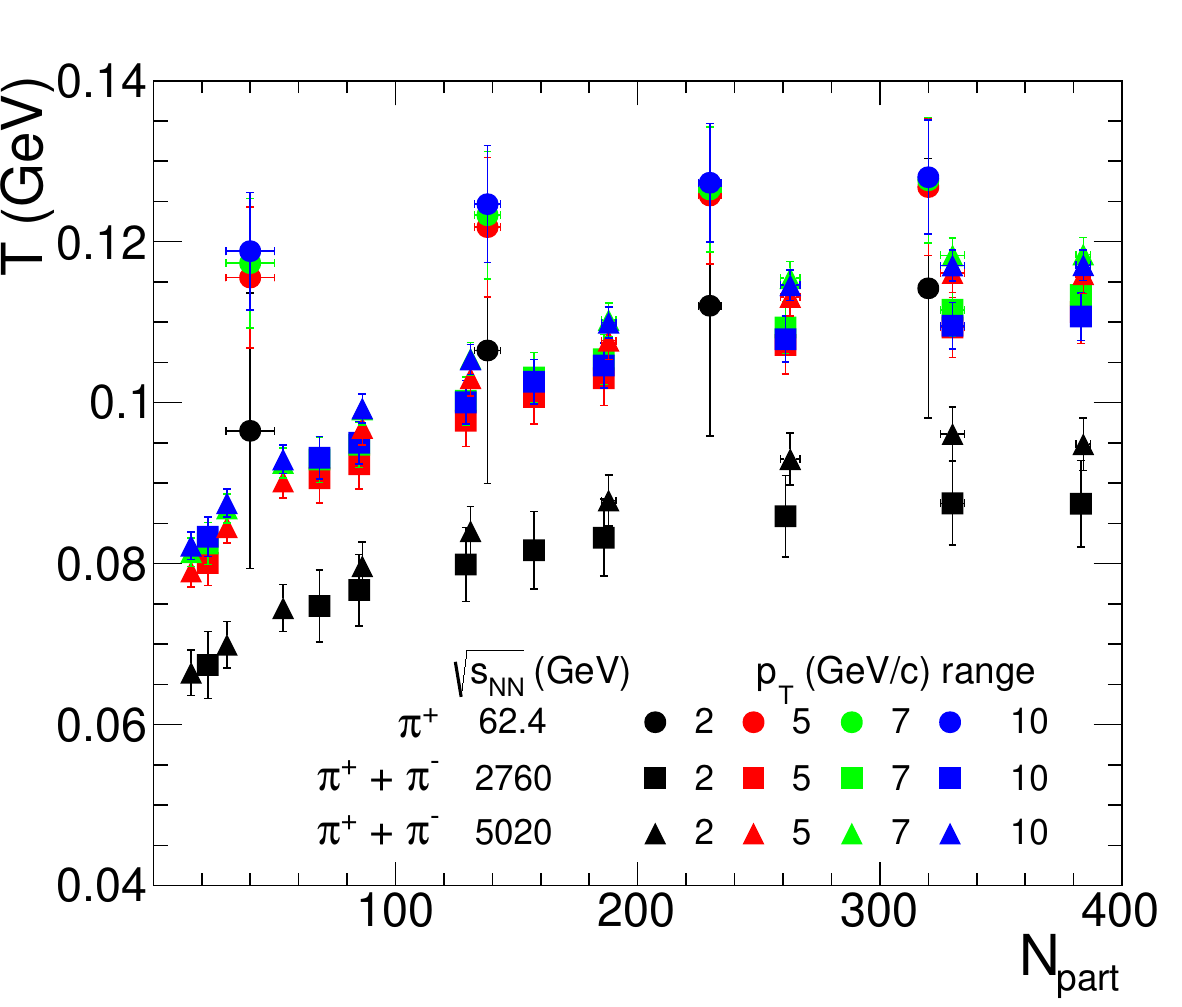}
	\caption{Tsallis parameters $q,~T$ as a function of centrality ($N_{\rm part}$) for the charged pions at different $p_{\rm T}$ ranges at $\sqrt{s_{NN}}=62.4, 2760, 5020$~GeV~\cite{STAR-62.4,ALICE-2.76-2016,ALICE-5.02-2020}.  The different markers and colours represent different $\sqrt{s_{NN}}$ and \pT~ranges, respectively.}
	\label{Fig:qT-Npart-2760}
\end{figure}

\begin{figure}[th!]
	\centering 
	\includegraphics[width=0.9\linewidth]{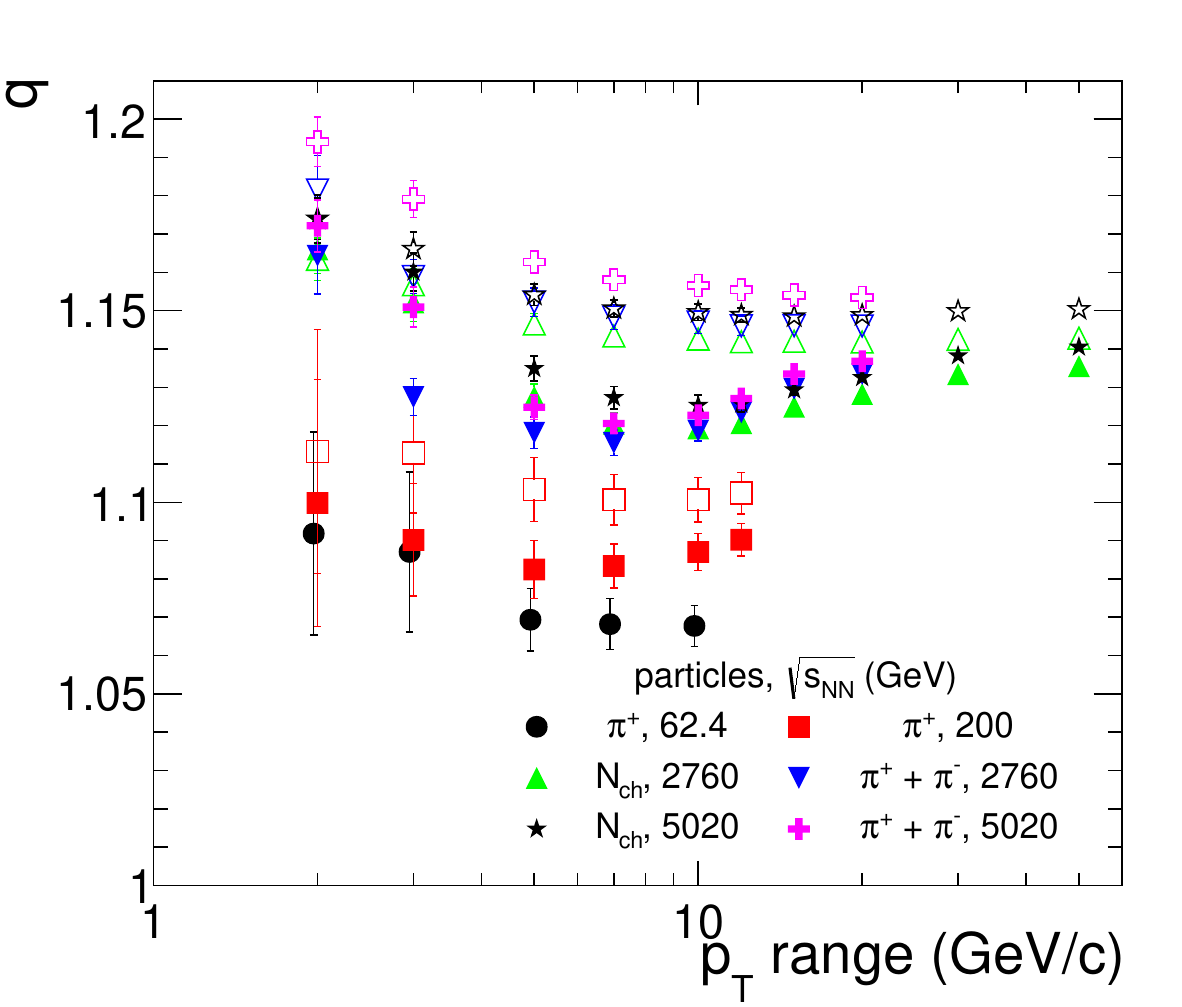}
	\includegraphics[width=0.9\linewidth]{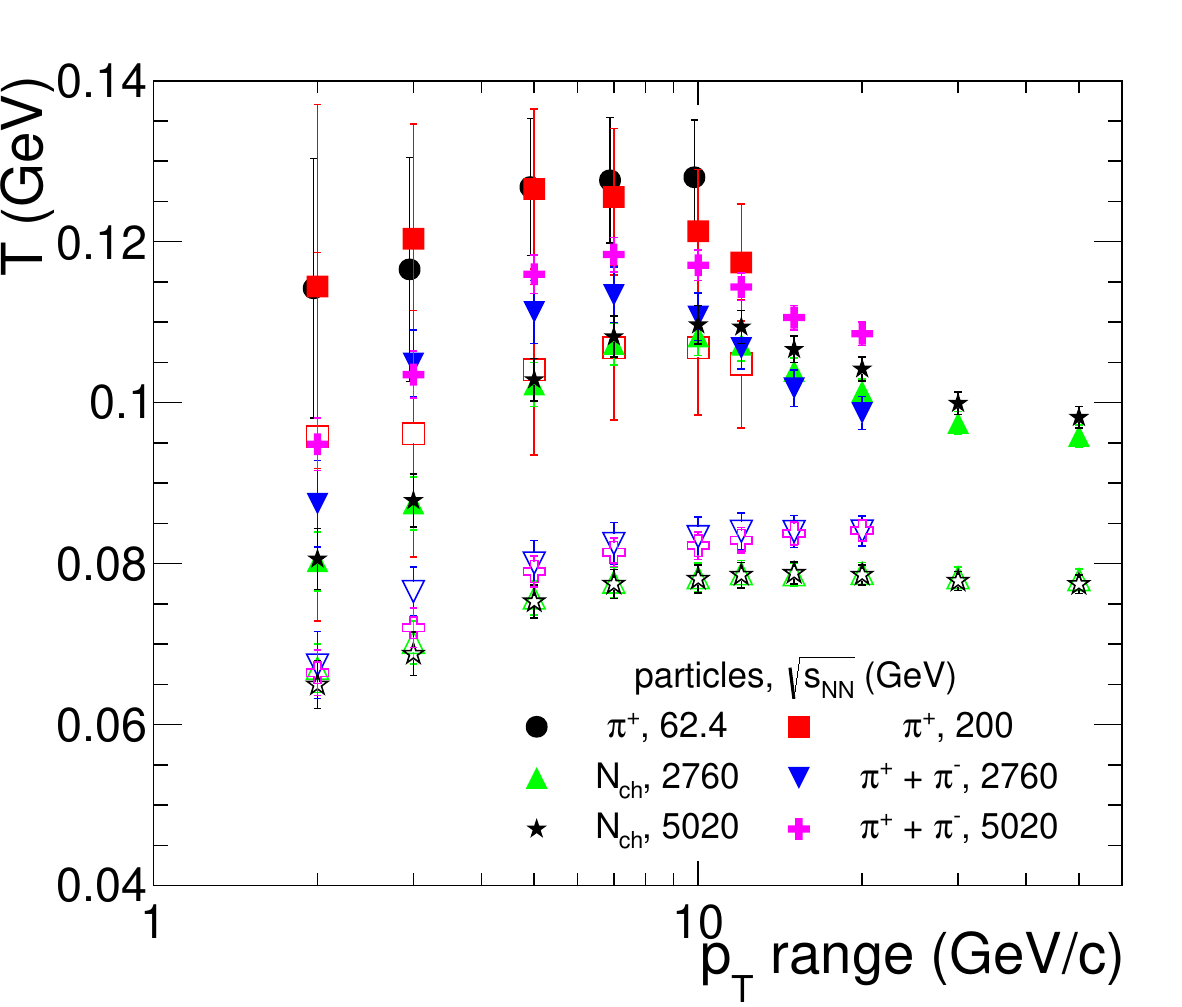}
	\caption{Tsallis parameters $q,~T$ as a function of $p_{\rm
            T}$ range for different collision energies. The parameters
          are at $\sqrt{s_{NN}}=$~62.4, 200, 2760, 5020~GeV for all
          charged particles $N_{\rm ch}$, all charged pions ($\pi^+ +
          \pi^-$) and $\pi^+$ as labeled by different
          markers~\cite{STAR-62.4,STAR-200,ALICE-2.76-2016,ALICE-2.76-Nch,ALICE-5.02-2020,ALICE-5.02-Nch}. 
           The
          open and solid markers correspond to 
          the parameters of the peripheral and central collisions, respectively.}
	\label{Fig:qT-pT}
\end{figure}

The Tsallis function has been shown to be successful in fitting $p_{\rm T}$ distributions over a broad range. In Fig.~\ref{Fig:qT-Npart-2760} we present the variation of the Tsallis fit parameters ($q$, and $T$) for charged pions as a function of centrality for different collision energies and for a set of fitting ranges ($0-2$ GeV/$c$, $0-5$ GeV/$c$, $0-7$ GeV/$c$, and $0-10$ GeV/$c$). In the figure, different energies are labeled by different markers whereas \pT~ranges are indicated using different colour schemes. We observe that:

 \begin{itemize}
 		
 \item {the fit parameters, $q$, $T$ depend on the fitting range in $p_{\rm T}$ for all centralities and collision energies. For the low-\pT~region, $q$ values have a big rise and $T$ values have a drop from RHIC to LHC energies. At mid-\pT~region ($5-10$ GeV/$c$), $q$ and $T$ are similar (within the uncertainties) at all energies.}

\item {for all centralities, $q$ values are different at low-$p_{\rm T}$ region($\sim$2~GeV/$c$), whereas, at mid-\pT~region (5--10 GeV/$c$) those are same within the uncertainties with higher values. In most of the cases, $q$ decreases from peripheral to central collisions for all $p_{\rm T}$ ranges. }

\item {the temperature, $T$, is almost constant at mid-\pT~region (5--10~GeV/$c$) with a higher value than low-\pT~fitting range ($\sim$2~GeV/$c$) for all collision energies. $T$ increases from peripheral to central collisions for all $p_{\rm T}$ ranges.}
\end{itemize}
The \pT~range dependencies of the Tsallis parameters of Au-Au collisions at 200~GeV data have similar behaviour as that of the Au-Au collisions at 62.4~GeV. Since the uncertainty on these datasets are large, only the Au-Au data at 62.4~GeV are plotted.

The behaviors of the fit parameters have been explored further by
plotting these as a function of the $p_{\rm T}$ ranges for different
collision energies. Depending on the availability of the experimental
data, the $p_{\rm T}$ ranges have been selected up to 2, 3, 5, 7, 10,
12, 15, 20, and 50~GeV/$c$. The evolution of the Tsallis parameters
with $p_{\rm T}$ range selections are shown in
Fig.~\ref{Fig:qT-pT}. The results shown in the figure correspond to
the fit values of all possible $p_{\rm T}$ spectra at
$\sqrt{s_{NN}}=$~62.4, 200, 2760 and 5020~GeV of all charged
particles, $\pi^+ + \pi^-$, and $\pi^+$. The open and solid markers
correspond to peripheral (70--80\%) and most central (0--10\%) collisions
respectively. We observe that, 
 \begin{itemize}
	\item {for peripheral collisions, $q$ initially decreases up to mid-\pT~region and then remains constant. For central collisions, $q$ decreases with an increase of $p_{\rm T}$ up to mid-\pT~region, 5--10~GeV/$c$, after which a slow rise is observed.}
	\item { for the peripheral collisions, $T$ increases slowly with \pT~and then remains constant. For central collisions, $T$ increases with $p_{\rm T}$ up to the mid-\pT~region, and then a slow decreasing trend is observed. }
\end{itemize}
The principal observation from the Fig.~\ref{Fig:qT-pT} is that depending on the collision energy and centrality, the \pT~spectra have different behaviour at different \pT~ranges. This is evident from the fact that the particle production mechanisms are different depending on the \pT~range probed. The charged pions observed at lower \pT~range are affected by resonance production, flow, coalescence, whereas at mid-\pT~region, contributions from hard pQCD processes and jets start to dominate.

\section{Discussion}\label{discussion}
\begin{figure}[th!]
	\centering 
	\includegraphics[width=0.9\linewidth]{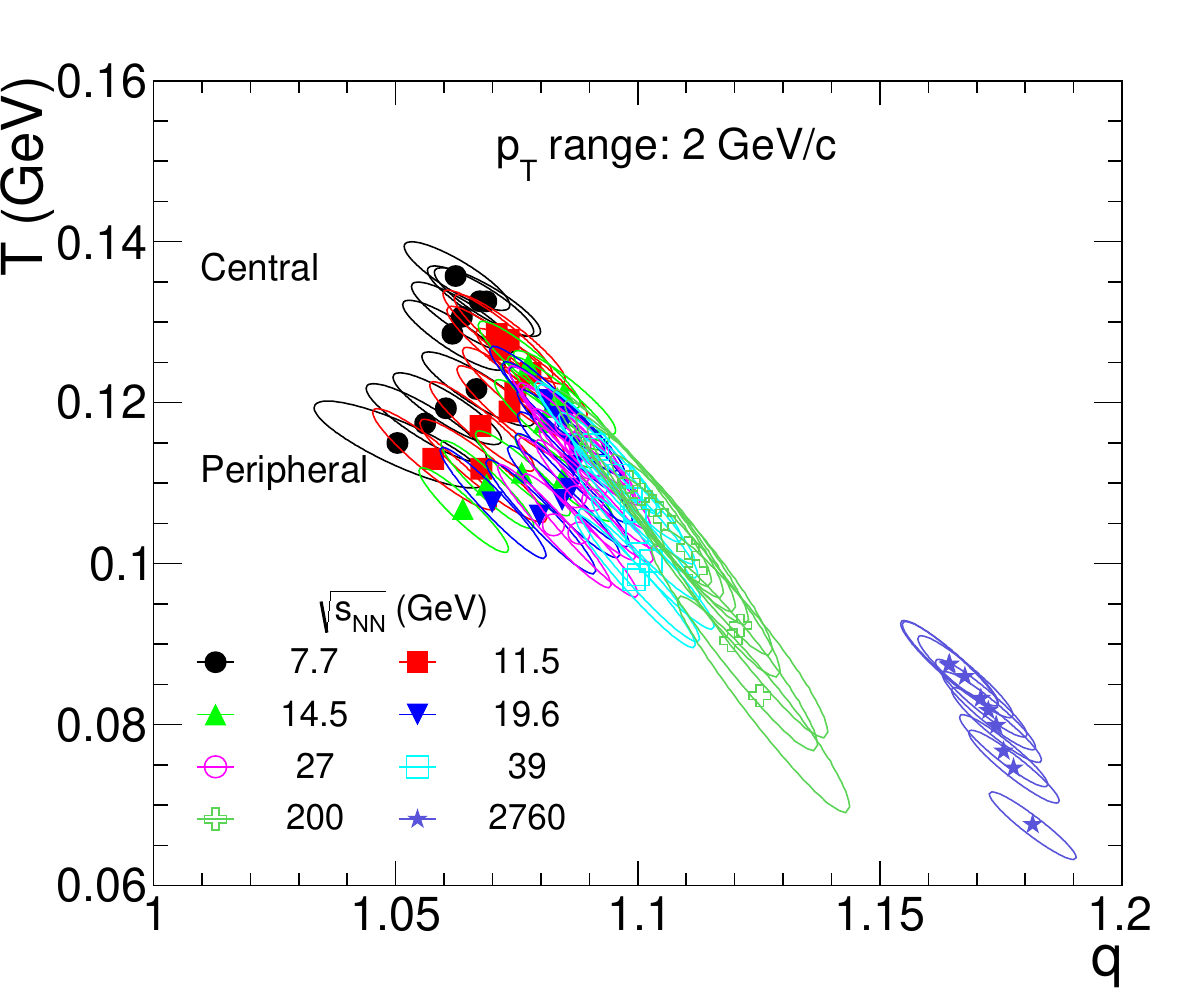}
	\includegraphics[width=0.9\linewidth]{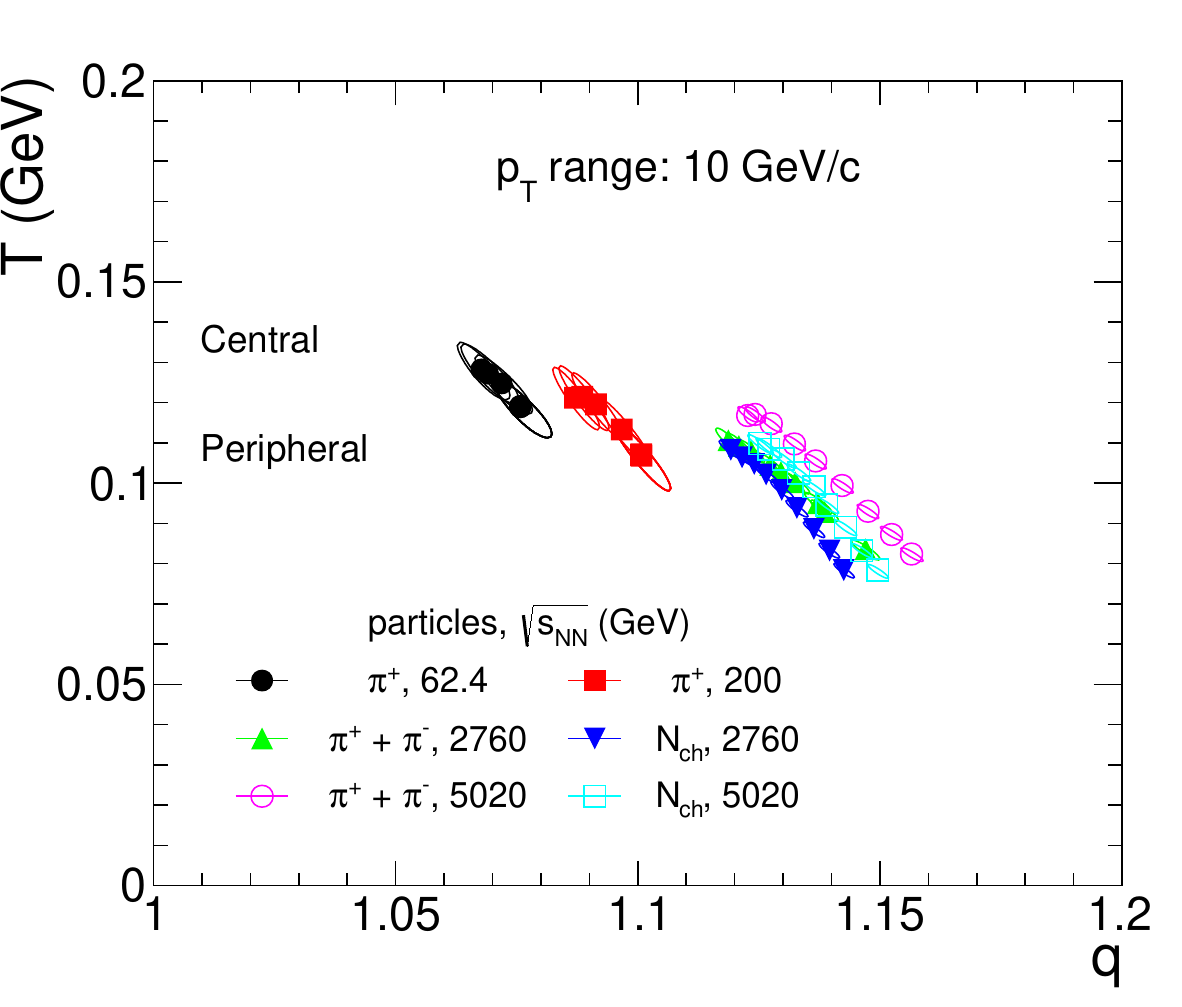}
	\caption{Profile plots of the temperature ($T$) and $q$ with one-$\sigma$ uncertainties ellipses for different centrality classes and collision energies for (upper panel) low $p_{\rm T}$
          range (up to 2~GeV/$c$) and (lower panel) $p_{\rm T}$ range up to
          10~GeV/$c$. 
	}
	\label{Fig:q-T_profile}
\end{figure}

Our study shows that the Tsallis parameters obtained by fitting the
$p_{\rm T}$ distributions vary with
respect to the fitting range in $p_{\rm T}$, centrality of the
collision and collision energy. The study of Tsallis parameters with
variation of centrality and collision energy as shown in
Fig.~\ref{Fig:q-Npart},~\ref{Fig:q-s} show that there may have
some correlation between the two parameters. 
This correlation between the
Tsallis fit parameters can be better understood by making profile
plots of $T$ and $q$, as shown in Fig.~\ref{Fig:q-T_profile}. In the profile plots the ellipses corresponding to the one-$\sigma$ uncertainties of the parameters $q$ and $T$ are shown. In the
upper panel of Fig~\ref{Fig:q-T_profile}, $T$ {\it vs.} $q$ are plotted
for fits up to $p_{\rm T}$ of 2~GeV/$c$ for pions for all
collision energies and collision centralities. 
We observe that for a
given $\sqrt{s_{NN}}$, central collisions have higher effective
temperatures compared to those of the peripheral collisions. In
addition, with increasing
collision energies $T$ decreases whereas $q$ increases as has been
found before.
In the lower panel of Fig~\ref{Fig:q-T_profile}, we plot the variation
of $T$ and $q$ for fit ranges up to $p_{\rm T}$ of 10~GeV/$c$ for
collision energies of 62.4~GeV and above. 
At the LHC energies, fit values from charged particles and identified charged pions are
plotted. The observed pattern is in general, similar
to what has been seen for fits with 
$p_{\rm T}$ up to 2~GeV/$c$, although the values of $q$ and
$T$ are different for the same collision energies. The uncertainties ellipses of the profile plots show that the parameters, $q$ and $T$, values have a mutual dependency and they have anti-correlation.   
From the lower panel of Fig~\ref{Fig:q-T_profile}, we observe that for the LHC energies the uncertainties ellipses are very small. Therefore, the Tsallis parameters are independent with precise values.  

We have mentioned before that  Boltzmann-Gibbs distribution with blast-wave
formula has been popularly followed to obtain the temperature at kinetic freeze-out
conditions in heavy-ion
collisions~\cite{STAR-BES1,STAR-14.5,ALICE-2.76-2013,ALICE-5.02-2020}. However,
it is limited to describe only small $p_{\rm T}$ ranges. From our study, it is observed that 
Tsallis distribution fits the spectra of long $p_{\rm T}$ ranges. 
The study of the $p_{\rm T}$ spectra of the identified particles in the
Tsallis framework shows that the effective temperature follows a
mass ordering with higher effective temperatures for particles
with larger masses as observed in other study~\cite{PHENIX-2004}. That can be seen from the flattening of the
spectra with higher mass particles. 
It is also observed that spectra of
peripheral collisions have better fitting compare to central
collisions. This is understood as the effective
temperatures from the Tsallis fit include the effect of radial flow.
These observations infer that accounting for the radial flow
part within the Tsallis distribution functions may better describe the spectra
for particles with higher masses and central collisions where
radial flow has significant contributions. The inclusion of radial
flow may describe the unusual values of $q$ and $T$ for the proton
spectra at the LHC energies.

\section{Summary} \label{summary}

In high-energy collisions, the transverse momentum spectra of produced
particles provide useful information regarding the
particle production mechanisms as well as freeze-out conditions of the
system. We have analyzed the \pT~spectra of identified and all charged
particles produced in Au--Au collisions at eight energies at RHIC and
Pb--Pb collisions at two energies at LHC using a the non-extensive
Tsallis statistics. The Tsallis function is found to be successful to
fit a wide range in \pT, with fit parameters, $q$, which is the
entropy index measuring the degree of non-additivity of the entropy of
the system, effective temperature, $T$, and a normalization parameter $V$ which
is proportional to the volume of the system.

For each collision centrality, with the increasing of the collision
energy, $q$ systematically increases and $T$
decreases. In the centrality dependency study we found that $q$ is
almost constant with an increasing trend for peripheral collisions in
LHC energy. The values of $T$ increase from peripheral to central
collisions. The parameter, $V$, increases monotonically with
centrality and collision energy. Dependence of the parameters, 
$q, T$, on the fitting range in \pT~has been observed, which also depends
on the collision energy. For central collisions the fit parameters are
found to have a strong dependence on the fitting ranges of 
$p_{\rm  T}$, however, for peripheral collisions there is a minor
dependency on $p_{\rm T}$. 
At mid-\pT~range, the Tsallis parameters are found to be constant within the uncertainty.
At low \pT~ranges, the particle production is
dominated by soft processes, whereas at mid \pT~ranges, pQCD processes
will have a larger effect. So, this observation may be interpreted as 
the effect of different physics
processes dominating different \pT~domains,  which also depend on
centrality. 
The profile plot of $T$ {\it vs.} $q$ with the
uncertainties ellipses shows an anti-correlation of the Tsallis parameters.

In literature, fits of $p_{\rm T}$ distributions with Tsallis function
are found to be successful in describing identified and all charged particles
in $pp$ collision system. In the present study, we have
found that in heavy-ion collision systems, the Tsallis distribution gives
good fits for $p_{\rm T}$ spectra of pions for full transverse momentum
ranges, but the fits are not as satisfactory for kaons and protons, 
particularly in central collisions. We have also found that the 
Tsallis distribution provides good fits for full $p_{\rm T}$ spectra in
case of all charged particles. This result is obvious as the pions are the most
abundant particles in all charged particles spectra. The centrality dependent
mass ordering in the effective temperature has been observed, which suggests
that the contribution of radial flow might need to be incorporated into the
standard Tsallis distribution function. Further investigation is
needed to include radial flow contributions in order to understand the full
nature of the evolving system and freeze-out in relativistic heavy-ion collisions.

\bigskip
%==================BIBLIOGRAPHY============================%
\bibliographystyle{unsrt}

\end{document}